\documentclass[11pt,oneside]{article}
\usepackage[english]{babel}
\usepackage{t1enc}
\usepackage{booktabs,amsthm, amsmath, amssymb, amsfonts, amsbsy, graphicx, epsfig, enumitem}
\usepackage{epstopdf}
\usepackage{soul}
\usepackage{parskip}
\usepackage{multirow}
\usepackage{mathrsfs}
\usepackage[colorlinks=true, pdfstartview=FitV, linkcolor=blue, citecolor=blue, urlcolor=blue]{hyperref} 
\usepackage[usenames]{color}
\usepackage[numbers]{natbib}
\usepackage{url}
\usepackage{accents,comment}

\usepackage{titlesec}

\titleformat{\subsubsection}[runin]
{\normalfont\large\bfseries}{\thesubsubsection}{0.7em}{}

\newtheorem{thm}{Theorem}[section]

\newtheorem{ex}[thm]{Example}
\usepackage[normalem]{ulem}

\usepackage[normalem]{ulem}

\begin{document}

\title{The least squares method for option pricing revisited}
\author{Maciej Klimek\footnote{Department of Mathematics, P.O.Box 480, Uppsala University, 751 06 Uppsala, Sweden; \emph{e-mail:} Maciej.Klimek@math.uu.se}\hspace{0.25cm}  and Marcin Pitera\footnote{Institute of Mathematics, Jagiellonian University, ul. {\L}ojasiewicza 6, 30-348 Krak{\'o}w. Poland; \emph{e-mail:} Marcin.Pitera@im.uj.edu.pl}}
\date{\today}

\maketitle

\begin{abstract}\noindent
It is shown that the the popular least squares method of option pricing converges even 
under very general assumptions. This  substantially increases the freedom of creating different implementations of the method, with varying levels of computational complexity and flexible approach to regression. It is also argued that in many practical applications even modest non-linear extensions of standard regression may produce satisfactory results. This claim is illustrated with examples.\\
\\
{\noindent \small
}
\end{abstract}

\section{Introduction}
For over a decade several variants of the so called least squares method of American option pricing have been widely used by financial practitioners and at the same time studied by researchers. The origins of the method can be found in the work of Carriere \cite{Car1996}, Tsitsiklis, Van Roy \cite{Tsi2001} (see also \cite{Tsi1999}), Longstaff, Schwartz \cite{LS} and Cl{\'e}ment, Lamberton, Protter \cite{Cle}. Basically, the method seeks a way of approximating conditional expectations needed in the valuation process either directly as in \cite{LS} and \cite{Cle}, or indirectly through the value function as in \cite{Tsi2001}.
A modification of the algorithm from \cite{LS} was studied in \cite{Cle} from the point of view of the convergence of the method. Subsequently, several papers on this subject have been published --- we will mention just a few of them related to the present article.

Glasserman and Yu \cite{Gla2004a} investigated in 2004 the convergence of the least squares like methods, where --- basically --- the necessary conditional expectations are approximated by finite linear combinations of approximating functions. More specifically, they looked into the problem of accuracy of estimations when the number of approximating functions and the number of simulated trajectories increase. They assumed that the underlying is a multidimensional Markov process. The rather pessimistic outcome, from the practical point of view, is that for polynomials as the approximating functions and for the conventional (resp. geometric) Brownian motion as the underlying, the number of required paths may grow exponentially in the degree (resp. the square of the degree) of the polynomials. Glasserman and Yu remarked that similar property may hold also for more general approximating functions (with the number of approximating functions replacing the maximal degree).
Also in 2004, Stentoft \cite{Ste2004} analyzed and extended the convergence results  presented in \cite{Cle}. In particular, he has considered the problem of choosing the optimal number of regressors in relation to the number of simulated trajectories. 
In 2005, Egloff \cite{Egl2005} proposed an extension to the original Longstaff-Schwartz \cite{LS} as well as Tsitsiklis-Van Roy (\cite{Tsi1999}, \cite{Tsi2001}) algorithms by treating the optimal stopping problem for multidimensional discrete time Markov processes as a generalized statistical learning problem. His results also improve those from \cite{Cle}. Egloff comments that despite very good performance of least squares algorithms in some practical calculations, precise estimates of the statistical quantities involved in these procedures may be difficult, leading to some less impressive performance in other cases.  
Zanger \cite{Zan2009} proposed in 2009 another extension to the least squares method by considering fairly arbitrary subsets of information spaces as the approximating sets. He has also produced some new and interesting convergence results showing in particular that sometimes the exponential dependence on the number of time steps can be avoided. It should be mentioned that the least squares approach can be also seen as part of the stochastic mesh framework proposed by Broadie and Glasserman (\cite{BroGla1997}, \cite{BroGla2002}; see also \cite{Liu2009} and \cite{Gla2010}). It should also be observed that two features seem to be common to the articles mentioned above. Firstly, the underlying is assumed to be Markovian. Secondly, the convergence rates of the method, in all its incarnations, are not encouraging from the computational point of view.

In the present paper, we extend the Cl{\'e}ment, Lamberton, Protter approach \cite{Cle} to 
 a fairly general setting for the regression approximating conditional expectations. Also, in a natural way,
the underlying does not have to be Markovian and the pay-offs are allowed to be path-dependent. 
While lack of Markov property can be easily circumnavigated in other ways, this always implies additional computational cost.
Obviously, by aiming at better approximation of conditional expectation, the potential computational complexity increases considerably. However, the main advantage of relaxation of the assumptions is the increase in freedom to customize the method. Moreover, we would like to argue that the least squares methods should be seen as a general framework leading to a variety of specific implementations. The main reason is essentially the fact that the information space for conditional expectation, or in other words its range, is in many interesting cases infinite dimensional. Inevitably, in these cases any approximation of conditional expectations, or value functions depending on conditional expectations, has to involve significantly restrictive extrinsic assumptions to make practical computations possible. While general convergence results are necessary to motivate the overall approach and some computational complexity may be addressed along the lines of \cite{Rus1997}, it is most likely that the future developments will evolve in the direction of simplified time-series models.  It is quite conceivable that an alternative source of realism and numerical efficiency could exploit the advances in both time-series analysis and frame theory (see e.g. \cite{K-M-O}). The empirical basis for such speculations comes from the fact that in many real problems even by taking only a few non-linear regressors, and sometimes ignoring lack of the Markov property, one might arrive at satisfactory results from the practical point of view. There seem to be much anecdotal evidence coming from the financial industry supporting the last statement and in this paper we provide further  corroborating evidence in the form of three empirical examples.

The material is organized as follows. The introduction is followed by a short review of consequences of the classic Dobrushin-Minlos theorem, which can lead to viable numerical approximations of conditional expectations. After recalling briefly how Snell envelopes are used in pricing of American-style options, we show that the methods proposed by Cl\'ement, Lamberton and Protter \cite{Cle} can be extended to cover the case of American style options  with a very general approach to regression. The setting includes  path dependent pay-offs and a non-Markovian multidimensional underlying. This is followed by three computational examples illustrating the viability of the method under rather restrictive assumptions. First, we present pricing of a one year Eurodollar American put and call options with different strike prices. Then, we use the least squares approach to price a 1.5 month American put option, whose payoff function depends on two market indices, namely DAX and EUROSTOXX50. Finally, we use the least squares algorithm to price two 1.5 month American put options, whose payoff function is based on a single market index under the assumption that the underlyings can be described by the Heston-Nandi GARCH(1,1) model ~\cite{HesNan2000}. Again, we will use  EUROSTOXX50 and DAX indices as the respective underlying instruments.

\section{Approximation of conditional expectation}
In this section we will introduce some basic notation and recall a classic result of Dobrushin and Minlos \cite{Dob}, which provides motivation, as well as a choice of practical recipes, for approximation of conditional expectations via the so called {\it admissible projection systems}.
The Dodrushin-Minlos theorem shows a specific example of such approximation, but of course there exist infinitely many non-polynomial constructions that would have the same property.

Let $(\Omega,\mathcal{F},\mathbb{P})$ be a probability space. Since we will be dealing only with random variables of finite variance, we can rely on the Hilbert space geometry in addressing the issues of interest (see \cite{Sma}). 
A closed subspace $S\subset L^2(\Omega,\mathcal{F},\mathbb{P})$ is said to be
\emph{probabilistic} if it contains constants and is closed with
respect to taking the maximum of two of its elements, i.e. if $X,Y\in S$, then $X\vee Y\in S$.
For any non-empty set $\mathbf{X}\subset L^2(\Omega,\mathcal{F},\mathbb{P})$, its
\emph{lattice envelope} $\mathrm{Latt}(\mathbf{X})$ is defined as the smallest probabilistic subspace of $L^2(\Omega,\mathcal{F},\mathbb{P})$ containing $\mathbf{X}$. Moreover, if $\mathbf{X}=\{X_1,\ldots,X_n\}$ and $\mathcal{B}_n$ denotes the $\sigma$-algebra of Borel sets in $\mathbb{R}^n$, then it is not difficult to prove that
$
\mathrm{Latt}(\mathbf{X})=
L^2(\Omega,\sigma(\mathbf{X}),\mathbb{P})=
L^2(\Omega,(X_1,\ldots,X_n)^{-1}(\mathcal{B}_n),\mathbb{P}).
$
The latter is sometimes referred to as the \emph{information space  generated by} $X_1,\ldots,X_n$.
Even if $\mathbf{X}$ consists of just one scalar random variable,  $\mathrm{Latt}(\mathbf{X})$ is typically infinite-dimensional. 
Since it is also the range of the orthogonal projection $\mathrm{E}[\cdot\,|\,X_1,\ldots,X_n]$, it would be desirable from the numerical standpoint to be able to approximate such projections with projections onto smaller finite-dimensional vector spaces using available least squares algorithms. However, approximating an orthogonal projection with infinite-dimensional range by projections onto finite dimensional subspaces makes most error estimates useless, unless the nature of the projected objects is somehow known beforehand. 

In order to construct finite-dimensional approximation of conditional expectation one could use the following theorem, which is a slight reformulation of a result of Dobrushin and Minlos
\cite{Dob}.
\bigskip
\begin{thm}\label{thm:DopMin} Let
$(\Omega,\mathcal{F},\mathbb{P})$ be a probability space and let
$\alpha>0$. 
Let $\mathcal{P}_n$ denote the space of all polynomials of $n$ real variables.
If $X_1,\ldots,X_n$ are random variables such that
$e^{|X_j|}\in L^\alpha(\Omega,\mathcal{F},\mathbb{P})$ for $j=1,\ldots,n$, then:
\begin{description}
\item[(a)] $P(X_1,\ldots,X_n)\in L^p(\Omega,\mathcal{F},\mathbb{P})$ for any
polynomial $P\in\mathcal{P}_n$ and $p\in[1,\infty)$;
\item[(b)] the vector space $\{P(X_1,\ldots,X_n)\,:\,P\in\mathcal{P}_n\}$
is dense in $L^p(\Omega,\sigma(X_1,\ldots,X_n),\mathbb{P})$ for
every $p\in[1,\infty)$.
\end{description}
\end{thm}

It should be noted that the converse to part (a) is false as shown in the following example.

\bigskip
\begin{ex}\emph{
Let $n=1$ and let $X_{1}$ be a discrete valued random variable with probability mass function 
\[
\mathbb{P}[X_{1}=m]=\frac{\frac{1}{m^{\ln m}}}{\sum_{m=1}^\infty\frac{1}{m^{\ln m}}},\quad m\in\mathbb{N}.
\]
Since  for any $q\geq 1$ and $\alpha>0$
\[
\sum_{m=1}^\infty\frac{m^q}{m^{\ln m}}<\infty
\textrm{ and }
\sum_{m=1}^\infty\frac{e^{\alpha m}}{m^{\ln m}}=\infty,
\]
the property {\bf (a)} from Theorem~\ref{thm:DopMin} is satisfied but $e^{|X_{1}|}\not\in L^{\alpha}(\Omega,\mathcal{F},\mathbb{P})$. $\blacksquare$}
\end{ex} 

If the probability measure $\mathbb{P}$ has a bounded support, in $\mathbb{R}^n$,
then the assumption of the Dobrushin-Minlos theorem is trivially satisfied. In fact, in this special case the conclusion of the theorem follows directly from the Stone-Weierstrass Theorem. It is also easy to see that
if $X$ is Gaussian, then $e^{|X|}\in L^1$. However, if $X$ is lognormal, then its moment generating function  does not exist in the interval $(0,\infty)$ and hence $e^{\alpha|X|}\not\in L^\alpha$ for all $\alpha>0$.

In concrete applications, the condition $e^{|X|}\in L^\alpha$  can sometimes
be achieved by changing the probability distribution of {\lq\lq}very large{\rq\rq} values of $|X|$. For instance, this can be accomplished by truncation of probability distribution or some direct attenuation of the random variable $X$. Another possibility is the use of suitable weight functions. In this context, the Dubrushin-Minlos theorem can be used to justify the density part in the construction of several classic polynomial bases in spaces of square integrable functions associated with the names of Jacobi,  Gagenbauer, Legendre, Chebyshev, Laguerre and Hermite (see e.g. \cite{Chi1978}).

Let $V$ be an information space generated by random variables $X_1,\ldots,X_n$. Suppose that one can furnish a sequence of Borel functions $q_m:\mathbb{R}^n\longrightarrow\mathbb{R}$, with $m\in\mathbb{N}$, such that the set $\{q_m(X_1,\ldots,X_n)\,:\,m\in\mathbb{N}\}$ is linearly dense in $V$ (e.g. with the help of the Dobrushin-Minlos theorem). Then the conditional expectation operator $\mathrm{E}[\cdot\,|\,X_1,\ldots,X_n]$ is the pointwise limit of the sequence of projections onto linear spaces $V^m=\{q_k(X_1,\ldots,X_n)\,:\,1\leq k\leq m\}$ as $m\nearrow\infty$. This observation leads to an auxiliary concept of admissible projection systems.

Given  a discrete time filtration $\{\emptyset,\Omega\}=\mathcal{F}_0\subset\mathcal{F}_1\subset\ldots
\subset\mathcal{F}_T\subset\mathcal{F}$ in the probability space $(\Omega,\mathcal{F},\mathbb{P})$, we define an \emph{admissible projection system} as a family
of orthogonal projections $P_t^m\,:\,L^2(\Omega,\mathcal{F},\mathbb{P})\longrightarrow L^2(\Omega,\mathcal{F},\mathbb{P})$, where 
$t=1,\ldots,T$ and $m\in\mathbb{N}$,
with ranges $V_t^1\subset V_t^2\subset V_t^3\subset\ldots\ $, whose union is dense in $L^2(\Omega,\mathcal{F}_t,\mathbb{P})$ for each value of $t$.

Note that for any such system and for any fixed $t$, we get pointwise convergence of the projections $P_t^m$ to $\mathrm{E}[\cdot\,|\,\mathcal{F}_t]$. However, this is not a norm convergence unless the underlying sequence of subspaces becomes constant after finitely many steps.

It is well known that Snell envelopes are useful in valuation of American put options in discrete time models (see e.g. \cite{Pli}, p.127). They also furnish the main theoretical ingredient of the least squares option pricing algorithm which is the main topic of this paper. The standard use of Snell envelopes can be easily extended to provide pricing algorithms for more general American style options, that is options that allow execution at any time prior to maturity, but with a wide variety of pay-off patterns. 

For a given probability space $(\Omega,\mathcal{F},\mathbb{P})$, let
$(\mathcal{F}_t)_{t=0}^T$ be a filtration, where
$\mathcal{F}_0=\{\emptyset,\Omega\}$ and $\mathcal{F}_T=\mathcal{F}$.
Assume that an adapted stochastic process $(Z_t)_{t=0}^T$ is integrable. One could look at $(Z_{t})$ as the intrinsic value process, that is the (discounted) value of executing some American-style option at time $t$. The \emph{Snell envelope of}  $(Z_t)$ is defined as the
adapted process $U_t$ such that $U_T=Z_T$ and $U_t=\max\left(Z_t,\mathrm{E}[U_{t+1}|\mathcal{F}_t]\right)$ for $t\in\{0,\ldots,T-1\}$.
The value $U_{0}$ corresponds to the the price of the option associated with pay-offs given by $(Z_{t})$. Indeed $(U_{t})$ can be seen as the application of dynamic programming principle to the optimal stopping problem $\sup\{\mathrm{E}Z_\nu:\nu\in\mathcal{C}_0^T\}$,
where $\mathcal{C}_0^T$ denote the set of all stopping times with values in the set $\{0,1,\ldots,T\}$ (cf. \cite{LamLap2007} and references therein, for basic properties of Snell envelopes and applications to pricing American-style options). The dynamic programming principle could be also rewritten in terms of the series of stopping times $(\tau_{t})$, defined recursively by putting $\tau_T=T$ and
\[
\tau_t=
t\mathbf{1}_{\{ Z_t\geq E[Z_{\tau_{t+1}}\,|\,\mathcal{F}_t]\}}
+
\tau_{t+1}\mathbf{1}_{\{ Z_t<E[Z_{\tau_{t+1}}\,|\,\mathcal{F}_t]\}},
\qquad t=1,\ldots,T-1.
\] 
In particular, we get $U_{t}=E[Z_{\tau_{t}}|\mathcal{F}_{t}]$ and consequently, $\tau_{0}$ is optimal for $(Z_{t})$.

The key element in any numerical implementation of Snell envelopes is the ability to approximate the conditional expectation operator.  Except for the finite case, one has to deal with infinite-dimensional spaces of random variables. Some elucidation seems to be in order here.
Given an admissible projection system $(P_{t}^{m})$, for a fixed $m\in\mathbb{N}$ we define the stopping times
$\tau_t^{[m]}$ by recursion, putting $\tau_T^{[m]}=T$ and 
\[
\tau_t^{[m]}=
t\mathbf{1}_{\{ Z_t\geq P_t^m(Z_{\tau_{t+1}^{[m]}})\}}
+
\tau_{t+1}^{[m]}\mathbf{1}_{\{ Z_t<P_t^m(Z_{\tau_{t+1}^{[m]}})\}},
\qquad t=1,\ldots,T-1.
\]

Then the following theorem generalizes a result due to Cl\'ement, Lamberton and  Protter (see Theorem 3.1 in \cite{Cle}):

\medskip
\begin{thm}\label{thm:1} 
If $(P_t^m)$ is an admissible
projection system, then
$
\lim_{m\to\infty}\mathrm{E}\left[Z_{\tau_t^{[m]}}\ |\ \mathcal{F}_t\right]
=\mathrm{E}[Z_{\tau_t}\ |\ \mathcal{F}_t]
$
for $t=1,\ldots,T$, where the convergence is in $L^2$. 
\end{thm}

\emph{Proof:} Despite a much more general setting we have adopted here, we can use standard properties of projections in Hilbert spaces and proceed as in \cite{Cle}. $\blacksquare$

Obviously, the above considerations remain valid for vector valued stochastic processes.

\section{The least squares method of option pricing}
Assuming that the filtration is generated by a discrete time multivariate stochastic process, we will show how to use Monte Carlo methods to approximate numerically the value for the optimal stopping for a given adapted process $(Z_{t})$, i.e. how to approximate the Snell envelope $(U_{t})$ of that process. To do so, given an admissible projection system, we basically need to approximate numerically $\mathrm{E}\left[Z_{\tau_t^{[m]}}\right]$ for $m\in\mathbb{N}$, due to Theorem~\ref{thm:1} and the fact that $U_0=\max(Z_0,\mathrm{E}[Z_{\tau_1}])$. 

In what follows we will denote the set of all real $(m\times n)$-matrices by $\mathbb{R}^{m\times n}$ with the convention that $\mathbb{R}^m=\mathbb{R}^{1\times m}$. Throughout the section we will use notation and methods similar to those introduced in \cite{Cle} but adapted to our less restrictive assumptions.

Suppose that $(X_t)_{t=0}^T$ is a discrete time $d$-dimensional stochastic process on the
probability space $(\Omega,\mathcal{F},\mathbb{P})$, with $X_0$ being a constant. This process is meant to represent the prices of the underlying assets for an American style option we wish to valuate. 
Let
$
X=(X_1,\ldots,X_T):\Omega\longrightarrow\mathbb{R}^{d\times T}
$
and let
$\mathcal{F}_t=\sigma\left(X_0,\ldots,X_t\right)=
\sigma\left(X_1,\ldots,X_t\right)$ for $t=1,\ldots,T$. Given a family of Borel functions
$
f_t:\mathbb{R}^{d\times (t+1)}\longrightarrow\mathbb{R}_+,$ where $t=0,\ldots,T,
$
we define $Z_t=f_t(X_0,\ldots,X_t)$ for $t=0,\ldots,T$.
This sequence represents suitably discounted intrinsic prices of the option we want to consider. Such a general choice of functions $f_t$ expands the potential applicability well beyond American put options.

Next, we need to chose an admissible projection system for the filtration associated with $X$. This is equivalent to choosing for each $t\in\{1,\ldots,T\}$ a suitable sequence of Borel functions
$
q_t^k:\mathbb{R}^{d\times T}\longrightarrow\mathbb{R}$,
where $k\in\mathbb{N},$
which depend only on the first $t$ column variables, and are such that the sequence $\{q_t^k(X)\}_{k\in\mathbb{N}}$ is linearly dense and linearly independent in the space $L^2(\Omega,\sigma(X_1,\ldots,X_t),\mathbb{P})$. Then, we can select an increasing sequence of integers $(k_m)_{m\in\mathbb{N}}$, such that the spaces
$
V_t^m=\mathrm{Lin}\{q_t^k(X)\,:\,k=1,\ldots,k_m\}
$
and the orthogonal projections $P_t^m:L^2(\Omega,\sigma(X),\mathbb{P})\longrightarrow V_t^m$ have all the right properties. The symbol {\lq\lq}$\mathrm{Lin}${\rq\rq} denotes the linear envelope of the given set of vectors.

If the stopping times $\tau^{[m]}$ are defined as in the previous section, then for some
$\alpha_t^m\in\mathbb{R}^{k_m\times 1}$
we have 
\[
P_t^m\left(Z_{\tau_{t+1}^{[m]}}\right)=e_t^m(X)\,\alpha_t^m,
\]
where the mapping $e_t^m$ is given by the formula
$
e_t^m=(q_t^1,\ldots,q_t^{k_m}):\mathbb{R}^{d\times T}\longrightarrow\mathbb{R}^{k_m}.
$
In view of our assumptions, the Gram matrix of the components of $e_t^m(X)$ (with respect to the inner product $(Y_1,Y_2)\mapsto\mathrm{E}[Y_1Y_2]$), that is the matrix 
$
A_t^m=\Big[
\mathrm{E}\left[q_t^i(X)q_t^j(X)\right]
\Big]_{1\leq i,j\leq k_m}
\in\mathbb{R}^{k_m\times k_m},
$
is invertible and hence
\[
\alpha_t^m=(A_t^m)^{-1}
\left[
\begin{array}{c}
\mathrm{E}\left[Z_{\tau_{t+1}^{[m]}}\, q_t^1(X)\right]\\
\vdots\\
\mathrm{E}\left[Z_{\tau_{t+1}^{[m]}}\, q_t^{k_m}(X)\right]
\end{array}
\right].
\]

Given a number $N$, the next step it to use Monte-Carlo simulation to generate independent trajectories
$
X^{(n)}=\left(X^{(n)}_1,\ldots,X^{(n)}_T\right)\in\mathbb{R}^{d\times T}
$
of the
process $X$, for $n=1,2,\ldots,N$. Each simulation has the fixed starting point  $X^{(n)}_0=X_0\in\mathbb{R}^{d\times 1}$.

Define
$
 Z_t^{(n)}:=f_t\left(X^{(n)}_0,\ldots,X^{(n)}_t\right)
$
and let
$ 
\widehat{Z}_t=\left[
Z^{(1)}_t,\ldots,Z^{(N)}_t
\right]^* \in\mathbb{R}^{N\times 1}.
$
This column vector consists simply of the values at time $t$ of all simulated trajectories of the process $Z$.
Define also
\[
V_t^{(m,N)} = \textrm{Lin} 
\left\{
\left[
\begin{array}{c}
q_t^{k}(X^{(1)})\\
\vdots\\
q_t^{k}(X^{(N)})
\end{array}
\right]\,:\,k=1,\ldots,k_m
\right\} \subset\mathbb{R}^{N\times 1}
\]
and
\[
P_t^{(m,N)}=\mathrm{Proj}_{V_t^{(m,N)}}:\mathbb{R}^{N\times 1}\longrightarrow\mathbb{R}^{N\times 1} 
\]
with respect to the inner product $\frac{\langle x,y\rangle}{N},$ where $\langle x,y\rangle$ denotes the standard scalar product.
Note that
\[
V_t^{(m,N)}=\mathrm{Lin}\left\{
\textrm{the columns of the matrix }
\left[
\begin{array}{c}
e_t^m(X^{(1)})\\
\vdots\\
e_t^m(X^{(N)})
\end{array}
\right]\in\mathbb{R}^{N\times k_m}
\right\}\subset\mathbb{R}^{N\times 1}.
\]

If we define the stopping times $\tau_t^{[m]}$ by letting $\tau_T^{[m]}=T$ and the formula
\[
\tau_t^{[m]}=
t\mathbf{1}_{\{ Z_t\geq P_t^m(Z_{\tau_{t+1}^{[m]}})\}}
+
\tau_{t+1}^{[m]}\mathbf{1}_{\{ Z_t<P_t^m(Z_{\tau_{t+1}^{[m]}})\}},
\qquad t=1,\ldots,T-1,
\]
then for some $\alpha_t^m\in\mathbb{R}^{k_m\times 1}$ we have
\[
P_t^m\left(Z_{\tau_{t+1}^{[m]}}\right)=e_t^m(X)\alpha_t^m.
\]
Similarly, if we define the approximative stopping times $\tau_t^{n,m,N}$ by requiring that 
$\tau_T^{n,m,N}=T$ and by putting
\begin{eqnarray*}
\tau_t^{n,m,N}&=&t \mathbf{1}_{ \left\{Z^{(n)}_{t}\geq
\pi_n\left[P_t^{(m,N)}(\widehat{Z}_{\tau_{t+1}^{n,m,N}})\right]\right\}
}+ \tau_{t+1}^{n,m,N} \mathbf{1}_{ \left\{Z^{(n)}_{t}<
\pi_n\left[P_t^{(m,N)}(\widehat{Z}_{\tau_{t+1}^{n,m,N}})\right]\right\}
},\\
&&\textrm{for }t=1,\ldots,T-1,
\end{eqnarray*}
where
$
\pi_n:\mathbb{R}^{N\times 1}\longrightarrow\mathbb{R}
$ is the projection on the $n$-th coordinate, 
then for some $\alpha_t^{(m,N)}\in\mathbb{R}^{k_m\times 1}$ we have
\begin{eqnarray*}
P_t^{(m,N)}\left( \left[
\begin{array}{c}
Z_{\tau_{t+1}^{1,m,N}}^{(1)}\\
\vdots\\
Z_{\tau_{t+1}^{N,m,N}}^{(N)}
\end{array}
\right] \right)&=& \left[
\begin{array}{c}
e_t^m(X^{(1)})\\
\vdots\\
e_t^m(X^{(N)})
\end{array}
\right] \alpha_t^{(m,N)}.
\end{eqnarray*}
Let $A_t^{(m,N)}$ denote the $(k_m\times k_m)$-Gram matrix associated with the columns
of the matrix
\[
\left[
\begin{array}{c}
e_t^m(X^{(1)})\\
\vdots\\
e_t^m(X^{(N)})
\end{array}
\right],
\]
(with respect to the inner product $\frac{\langle x,y\rangle}{N}$).
Then this is simply the Gram matrix estimator for the given sample. 

Then $\alpha_t^{(m,N)}$ is a solution of the equation
\[
A_t^{(m,N)}\alpha_t^{(m,N)}=
\frac{1}{N}
\left[
\begin{array}{c}
e_t^m(X^{(1)})\\
\vdots\\
e_t^m(X^{(N)})
\end{array}
\right]^*
\left[
\begin{array}{c}
Z^{(1)}_{\tau_{t+1}^{1,m,N}}\\
\vdots\\
Z^{(N)}_{\tau_{t+1}^{N,m,N}}
\end{array}
\right].
\]
By the Law of Large Numbers
$A_t^{(m,N)}\stackrel{a.s.}{\longrightarrow}A_t^m$ as $N\to\infty$,
and
hence for sufficiently large $N$ the matrix $A_t^{(m,N)}$ is
invertible (almost surely). In this case
\[
\alpha_t^{(m,N)}=
\frac{1}{N}
\left(
A_t^{(m,N)}
\right)^{-1}
\left[
\begin{array}{c}
e_t^m(X^{(1)})\\
\vdots\\
e_t^m(X^{(N)})
\end{array}
\right]^*
\left[
\begin{array}{c}
Z^{(1)}_{\tau_{t+1}^{1,m,N}}\\
\vdots\\
Z^{(N)}_{\tau_{t+1}^{N,m,N}}
\end{array}
\right].
\]
For convenience we will write
$
\alpha^{m}=\left(\alpha_1^{m},\ldots,\alpha_{T-1}^{m}\right)
$ and 
$
\alpha^{(m,N)}=\left(\alpha_1^{(m,N)},\ldots,\alpha_{T-1}^{(m,N)}\right).
$
Both objects are $k_m\times(T-1)$-matrices.

The next theorem is a direct extension of Theorem 3.2 and Lemma 3.2 from \cite{Cle}.

\medskip\begin{thm} With the above notation,  as  $N\to\infty$,
\[
\frac{1}{N}\sum_{n=1}^N
Z^{(n)}_{\tau_{t}^{n,m,N}}\stackrel{a.s.}{\longrightarrow}
\mathrm{E}\left[Z_{\tau_t^{[m]}}\right],\qquad t=1,\ldots,T.
\]
\label{thm:2}
\end{thm}
\emph{Proof:} Define
$
B_t=\{
(a^m,z,x)\,: z_t<e_t^m(x)a_t^m
\}\subset
\mathbb{R}^{k_m\times(T-1)}\times\mathbb{R}^{T}\times\mathbb{R}^{d\times T}
$
for $t=1,\ldots,T-1,$ where $a^m=(a_1^m,\ldots,a_{T-1}^m)$, $z=(z_1,\ldots,z_T)$,
and $x=(x_1,\ldots,x_T)$. By $B_t^c$ we will denote the complement of $B_t$.
We define an auxiliary function
$
F_t:
\mathbb{R}^{k_m\times(T-1)}\times\mathbb{R}^{T}\times\mathbb{R}^{d\times T}
\longrightarrow\mathbb{R},
$
by recursion by putting $F_T(a^m,z,x)=z_T$ and $F_t(a^m,z,x)=
z_t\mathbf{1}_{B_t^c}+F_{t+1}(a^m,z,x)\mathbf{1}_{B_t}$ for $t=1,\ldots,T-1.$
It is easy to see that
\[
F_t(a^m,z,x)=
z_t\mathbf{1}_{B_t^c}+
\sum_{s=t+1}^{T-1}z_s\mathbf{1}_{B_t\cap\ldots\cap B_{s-1}\cap B_s^c}+
z_T\mathbf{1}_{B_t\cap\ldots\cap B_{T-1}}
\]
for $ t=1,\ldots,T-1$.
Moreover,
$F_t(a^m,z,x)$ is independent of $a_1^m,\ldots,a_{t-1}^m$, $F_t(\alpha^m,Z,X)=Z_{\tau_t^{[m]}}$ and $F_t(\alpha^{(m,N)},Z^{(n)},X^{(n)})=Z^{(n)}_{\tau_t^{n,m,N}}.$
For $t=2,\ldots,T$ define also two other auxiliary functions $
G_t(a^m,z,x)=F_t(a^m,z,x)e^m_{t-1}(x)$ and $\psi_t(a^m)=\mathrm{E}[G_t(a^m,Z,X)]$.
Using this notation, one can see that for $t=1,\ldots,T-1$:
\begin{align}
\alpha_t^m &= (A_t^m)^{-1}\psi_{t+1}(\alpha^m); \label{alpha}\\
\alpha_t^{(m,N)} &= (A_t^{(m,N)})^{-1}
\frac{1}{N}\sum_{n=1}^NG_{t+1}(\alpha^{(m,N)},Z^{(n)},X^{(n)}).
\label{alphaN}
\end{align}
The following estimate is a higher-dimensional counterpart of Lemma 3.1 in \cite{Cle}
and can be derived along the same lines as that lemma:
\begin{equation}
|F_t(a,z,x)-F_t(\tilde{a},z,x)|\leq\sum_{s=t}^T|z_s|
\left[
\sum_{s=t}^{T-1}
\mathbf{1}_{\{|z_s-e^m_s(x)\tilde{a}_s|\leq|e^m_s(x)|\|\tilde{a}_s-a_s\|\}}
\right],
\label{F-est}
\end{equation}
where $1\leq t\leq T-1$,
$a=(a_1,\ldots,a_{T-1})\in\mathbb{R}^{k_m\times(T-1)}$, $\tilde{a}=(\tilde{a}_1,\ldots,\tilde{a}_{T-1})\in\mathbb{R}^{k_m\times(T-1)}$, $z\in\mathbb{R}^{T}$ and $x\in\mathbb{R}^{d\times T}$.

Using (\ref{alpha}, \ref{alphaN}, \ref{F-est}), and under the technical assumption that  $\mathbb{P}(e_t^m(X)\alpha_t^m=Z_t)=0$,
the reasoning from \cite{Cle} can be easily modified to  work within our more general setup. In general, this additional technical requirement can be fulfilled by using approximation of the contract functions $f_t$ by functions with probabilistically {\lq\lq}negligible{\rq\rq} fibers and by introduction of small amount of random noise perturbing the probability distribution of $X_t$. $\blacksquare$ 

\bigskip
Theorems \ref{thm:1} and \ref{thm:2} provide a recipe for approximation of $\mathrm{E}[Z_{\tau_1}]$ and hence also $U_0=\max\left(Z_0,\mathrm{E}[Z_{\tau_1}]\right)$, as required.

\section{Examples}
In this section we show three examples of applications of the above least squares algorithm. The first example covers American call and put options written on Eurodollar futures, which are assumed to conform to the Brace-Gatarek-Musiela model~\cite{BraGatMus1997}. Next we price basket and dual-strike American put options for EUROSTOXX50 and DAX indices, under the standard bivariate Brownian dynamics. Finally, we show how to price univariate American put options, both for EUROSTOXX50 and DAX indices, assuming that the dynamics of the underlyings could be expressed using the Heston-Nandi GARCH(1,1) model~\cite{HesNan2000}.

We have decided not to include convergence speed analysis, as it would make the presented examples much more complicated (e.g. the proper variance reduction technique is a crucial step for any market implementation), without adding much to the conclusions drawn in this paper. We refer to~\cite{Cle,BevJos2009}, and references therein, for detailed analysis about convergence speed in univariate Markovian case. For transparency, we use only the standard models for parameter estimation and Monte Carlo simulation. In particular, only the prices of the underlyings are used for calibration purposes and no Monte Carlo variance reduction technique is implemented. Nevertheless, we present the (smoothed) density function of simulated prices for every example (see Figures~\ref{F:MCprice},~\ref{F:optionBASKET}~and~\ref{F:optionHN}) to give some insight into the accuracy of our implementation. It should be noted that while our examples are rather straightforward, the accuracy seems to be satisfactory. This allow us to be optimistic about the least squares algorithm approach for option pricing, even when the dynamics of the underlyings is complicated and no theoretical price is known.

Our implementation of the least squares algorithm is based on the in-the-money realizations to speed up the convergence and reduce the number of polynomials needed to achieve sufficient level of accuracy. It is worth mentioning that in real-world models, to improve the convergence rate and the speed of the algorithm one might use additional information available from the market (e.g. prices of various derivatives, based on the same underlying instruments) as well as various modifications of the standard Monte Carlo algorithm (for more advanced models cf. \cite{BevJos2009} or \cite{Gla2010} and references therein).

All computations were done using {\bf R 2.15.2} (64-bit). In particular we have used the libraries {\bf fOptions} (for Heston-Nandi parameter calibration, CRR prices and Monte Carlo simulation), {\bf orthopolynom} (for different base functions in L-S algorithm), {\bf timeSeries} (for market data handling) and {\bf Rsge} (for parallel computations).

\subsection{Eurodollar options}
In this subsection we use the least squares algorithm to price one year Eurodollar American put and call options with different strike prices, given the real-market daily prices of the Eurodollar futures. It should be noted that the standard Black-Scholes model cannot be used when the option price is based on more than one LIBOR rate (e.g. when the option's lifetime is longer than 3 months). This is due to the fact that forward rates over consecutive time intervals are related to each other and cannot all be log-normal under the same spot risk-neutral measure. Consequently, models of such instruments in the standard risk-neutral setting are based on non-Markovian dynamics. A.~Brace, D.~Gatarek and M.~Musiela~\cite{BraGatMus1997} proposed a model which can overcome this inconvenience (BGM Model) by utilizing a forward arbitrage-free risk-neutral measure. In the literature, it is also referred to as the LIBOR Market Model (LMM). It is worth mentioning that the dynamics of interest rates described in BGM model is very closely related to the Heath-Jarrow-Morton (HJM) Model. Next we will present a brief overview of the BGM model, followed by some basic information concerning the setup of the least squares algorithm.

\subsubsection{The Brace-Gatarek-Musiela Model.}
The Brace-Gatarek-Musiela model is a stochastic model of time-evolution of interest rates. It will be used here to simulate the (Monte Carlo) paths of LIBOR futures. We will now outline a simplified version of the model that suits our framework, and we will make some comments on the estimation procedure. Let $T_{0}=0$ and $T_{i}=T_{i-1}+\frac{3}{12}$ for $i=1,2,3,4$. In reality the dates of expiration for the consecutive Eurodollar futures differ slightly from 90 days. This might  potentially have an impact on the results, especially when we consider short term options. Nevertheless, we will use the theoretical values for simplicity. Let $L_{0}$ be a spot LIBOR rate and let $L_{i}:[0,T_{i}]\times\Omega\rightarrow \mathbb{R}$ be the $i$-th forward LIBOR rate. Assuming $d$ sources of randomness, the dynamics of the $i$-th LIBOR rate can be described by the equation
$$d\log L_{i}(t)=\left(\sum_{j=i(t)}^{i}\frac{\delta_{j}L_{j}(t)}{1+\delta_{j}L_{j}(t)}\sigma_{j}(t)-\frac{\sigma_{i}(t)}{2}\right)\sigma_{i}(t)dt +\sigma_{i}(t)dW^{\mathbb{Q}_{\textrm{Spot}}}(t),$$
where $t\in[0,T_{i}]$, $\delta_{i}=T_{i+1}-T_{i}=3/12$ is the length of the accrual period of the $i$-th LIBOR forward rate, $\sigma_{i}(t):[0,T_{i}]\times \Omega\rightarrow \mathbb{R}^{d}$ is the instantaneous volatility of the $i$-th LIBOR forward rate, $i(t)$ denotes the index of the bond (corresponding to the appropriate Eurodollar future) which is first to expire at time $t$, and finally, $W^{\mathbb{Q}_{\textrm{Spot}}}(t)$ is a standard ($d$-dimensional) Brownian motion under the spot LIBOR measure $\mathbb{Q}_{\textrm{Spot}}$ (see~\cite{Jam1997} for more details). We are assuming here that the sources of randomness are independent of each other and that the proper dependency structure is modelled with $\sigma_{i}$. For the Monte Carlo simulation we will use a standard Euler discretization of the above SDE, with the time step $\Delta t =\frac{1}{360}$, i.e. 
\begin{equation}\label{BGMdynamics}
\Delta\log L_{i}(t)=\left( \sum_{j=i(t)}^{i}\frac{\delta_{j}L_{j}(t)}{1+\delta_{j}L_{j}(t)}\sigma_{j}(t)-\frac{\sigma_{i}(t)}{2}\right)\sigma_{i}(t)\Delta t +\sigma_{i}(t)\epsilon_{t}\sqrt{\Delta t},
\end{equation}
where $\epsilon_{t}\sim \mathcal{N}(0,\mathbf{I})$ is a $d$-dimensional standard normally distributed random vector. In our implementation we will use $d=3$. To calibrate the model we need to define the functions $\sigma_{i}(t)$, for $i=1,2,3,4$. We will assume that $\sigma_{i}(t)$ (for $i=1,2,3,4$) is time homogeneous, i.e., that there exists a function $\lambda=(\lambda_1,\lambda_2,\lambda_3): [0,T]\rightarrow\mathbb{R}^{3}$ such that
$\sigma_{i}(t)=\lambda(T_{i}-t)$ for $t\in[0,T_{i}]$ and $i=1,2,3,4$. We will provide values of $\lambda(T_i)$ for $i=1,2,3,4$ and assume that $\lambda(t)=\lambda(T_{i})$ for $t\in [T_{i-1},T_{i}]$.

We will apply the Principal Components Analysis (PCA) to Eurodollar futures data to approximate the values of the $4\times 3$ matrix $\Lambda=[\lambda_j(T_i)]$.
In other words, we base our estimation process on the correlation between the Eurodollar futures. The difficulty with calibration of PCA is that Eurodollar futures have fixed maturity dates, and so for a given $T$ we can monitor a contract with volatility $\lambda(T)$ only once per three months. To overcome this, we will use linear interpolation of the quoted prices of Eurodollar futures (which is in fact a common market practice). It should be noted that we need the $L_{5}$ prices to perform such interpolation. Using that approach to Eurodollar futures prices, we obtain the values of contracts with volatility $\lambda(T_i)$ (i=1,2,3,4) for every trading day $t$. We also use linear interpolation of forward LIBOR rates for the days when the market is not operating (i.e. 
we interpolate the contract prices using known quotes from the last trading day before and the next trading day after the date in question). 
Because of that assumption, in order to conduct the PCA and to estimate $\sigma_{i}$ (for $i=1,2,3,4$), we will need (for each day) the prices of the five Eurodollar futures closest to delivery. Let us now comment on the PCA estimation process. We assume that
$$\lambda_j(T_i)=\frac{\Theta_{i}s_{j}\alpha_{i,j}}{\sqrt{\sum_{k=1}^{d}s_{k}^{2}\alpha_{i,k}^{2}}},$$
for $i=1,2,3,4$ and $j=1,2,3$. Here, $s_{j}^{2}$ denotes the variance of the $j$-th factor computed by PCA (with $s_{1}^{2}\geq s_{2}^{2}\geq s_{3}^{2}$), $\alpha_{i,j}$ measures the influence of the $j$-th factor when the time to maturity is in the period $[T_{i-1},T_{i}]$ and $\Theta_{i}:=\sum_{j=1}^{3}s_{j}\alpha_{i,j}$ is the total volatility in the $i$-th period. We also assume that the factors are uncorrelated  and that the relative influence of every factor is 1 (i.e. for $j_{1},j_{2}\in\{1,2,3\}$ we have  $\sum_{i}^{4}\alpha_{i,j_1}\alpha_{i,j_2}=0$ if $j_{1}\neq j_{2}$ and $\sum_{i}^{4}\alpha_{i,j_1}\alpha_{i,j_2}=1$ if $j_{1}=j_{2}$). Combining (\ref{BGMdynamics}) and the parameters from the PCA we will be able simulate Eurodollar futures paths.

\subsubsection{Setup, data details and the least squares method parameters.}
We wish to price the quarterly Eurodollar American call and put options EDZ2 (GEZ2 in Globex notation; it means that the underlying instrument is the December 2012 Eurodollar future). The first trade day for EDZ2 is December 13, 2010, and the expiration date is December 17, 2012. We will estimate the value of several such put and call options during the period from December 20, 2011 to January 20, 2012, with different strike prices - ranging from 98.00 to 99.75. While the values of the American call options could be computed without the use of the least squares algorithm, because they coincide with the European calls, we will calculate them anyway to provide more insight into how the parameters are fitted to the market data. In other words, we wish to check empirically if the differences between the market prices and the computed prices are the result of badly fitted model parameters or are due to a problem with accuracy of the least squares algorithm.

For the calibration purposes we will use the daily closing prices of Eurodollar futures and the spot LIBOR rate. Given a date $t$, we will use a period of the same length as the time to maturity of the option (i.e. if the option lifetime is 300 days, then we take last 300 days data before time $t$ to calibrate our model).

The least squares algorithm needs several inputs. As the functions generating the {\lq\lq}information about the past{\rq\rq} we use the standard exponentially weighted Laguerre polynomials of degree not greater than 3. 
Our implementation is based on the  Monte Carlo simulation of the $L_{4}$ values obtained using \eqref{BGMdynamics}. The algorithm needs also formulas for the interest rate (for the purpose of discounting) in two instances. Firstly, to discount the values of options from one period to another (in the recursive step-by-step part). Secondly, to compute the final price of the option (i.e. to discount the optimal prices from every simulation to time $T_0=0$). While the second interest rate could be associated with standard spot LIBOR rate, the first one must be based on the evolution of assets (i.e. for every path in the Monte Carlo run, one must estimate separately the spot rate at time $t$  using the prices of Eurodollar contracts).

\subsubsection{Estimation and numerical results.}
\begin{figure}[!ht]
\begin{center}
\includegraphics[width=6cm]{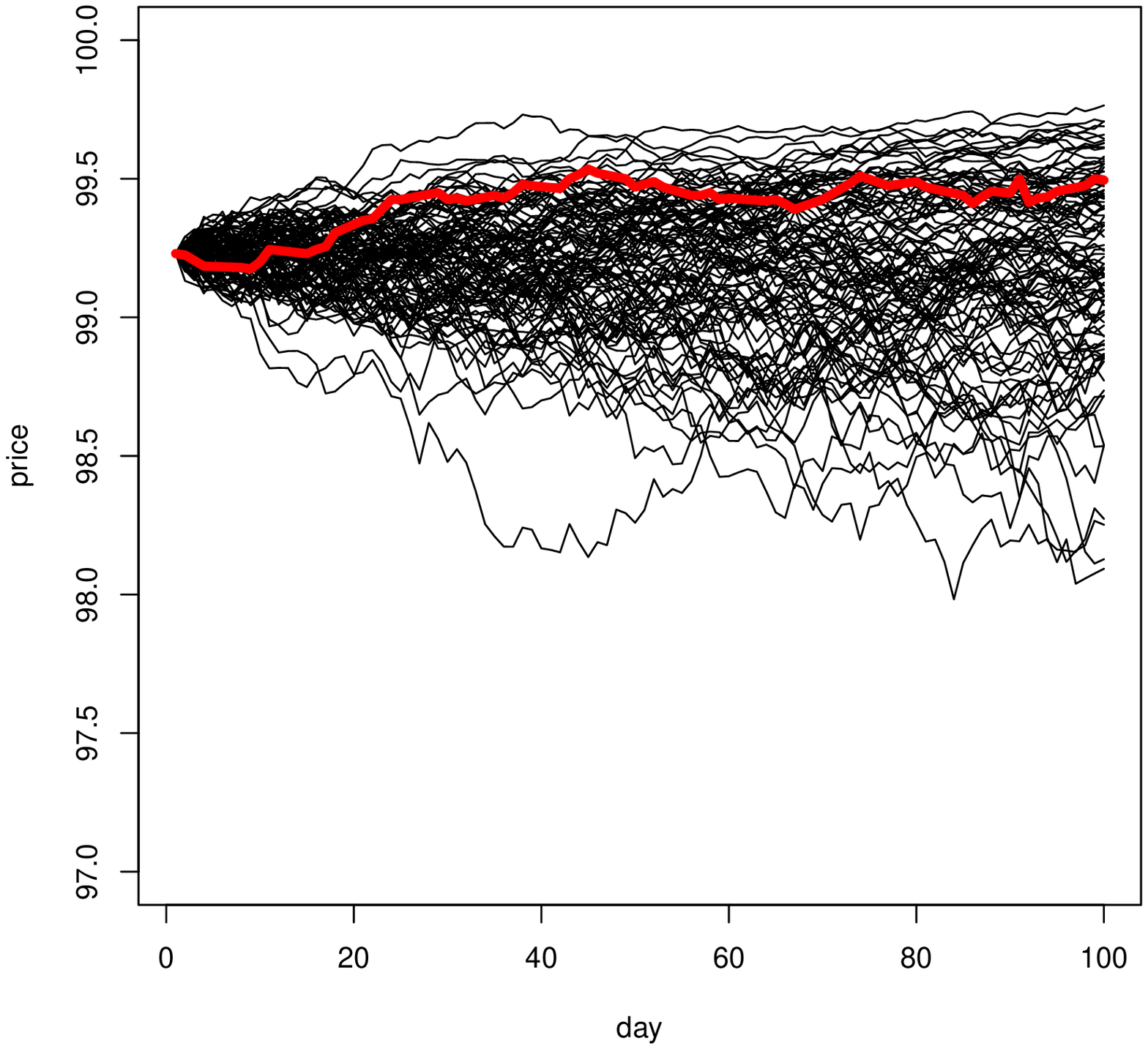}
\includegraphics[width=6cm]{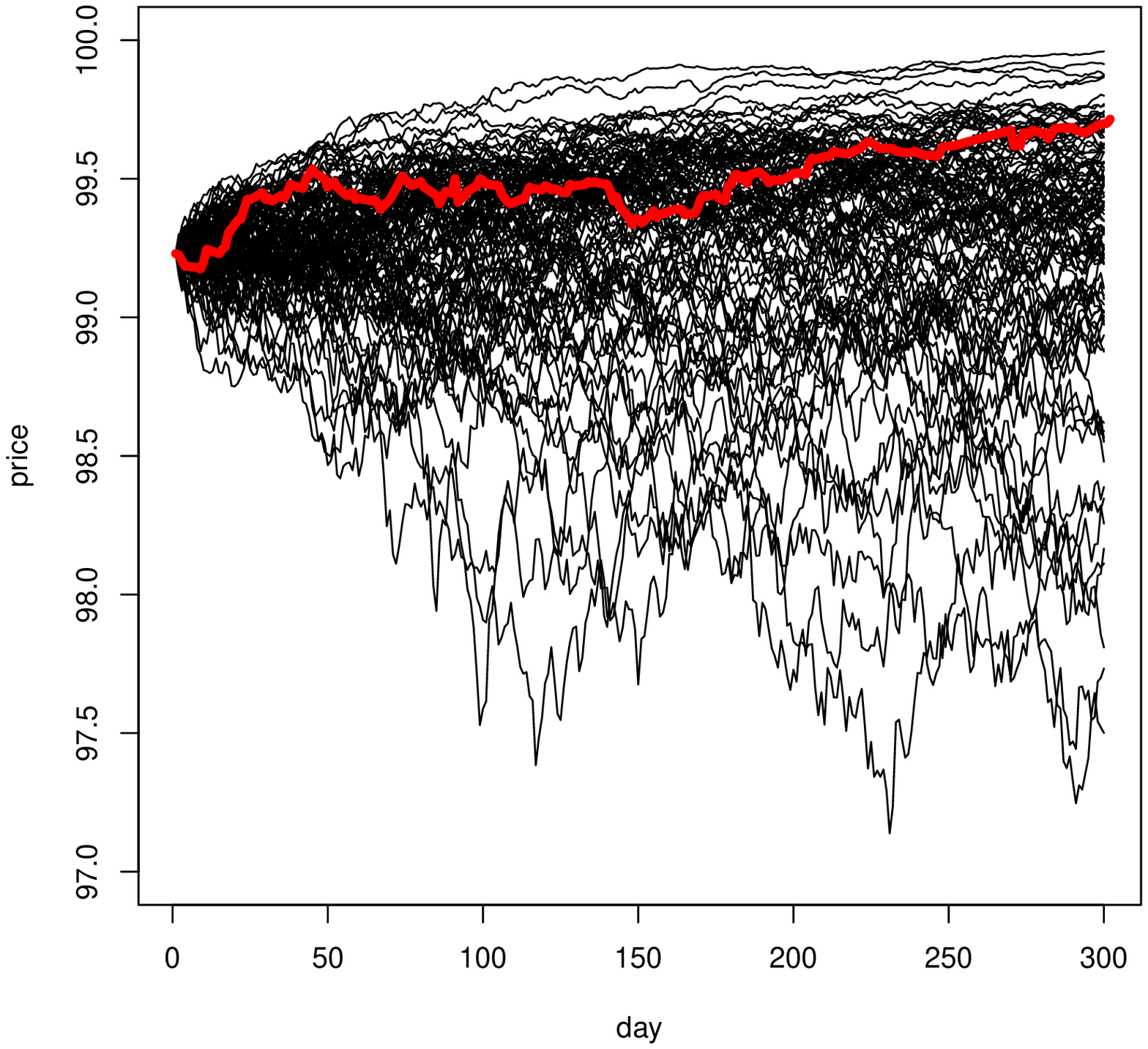}
\end{center}
\vspace{-20pt}
\caption{Examples of 100 Monte Carlo paths for the $L_{4}$ contact (for December 20, 2011) and the realized path (red) during the first 100 and 300 days.} \label{F:MCexample}
\end{figure}

\begin{figure}[!ht]
\begin{center}
\includegraphics[width=6cm]{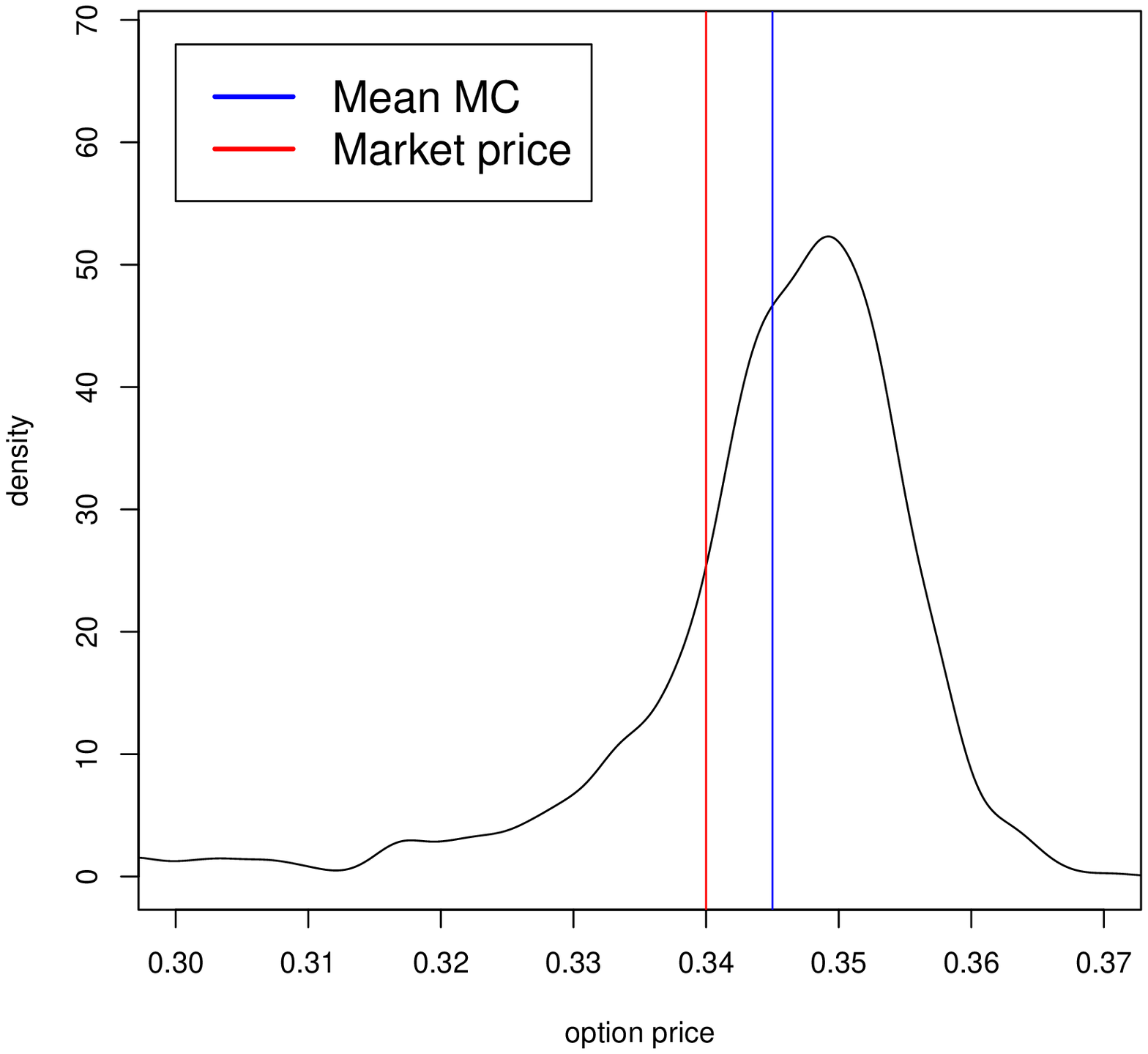}
\includegraphics[width=6cm]{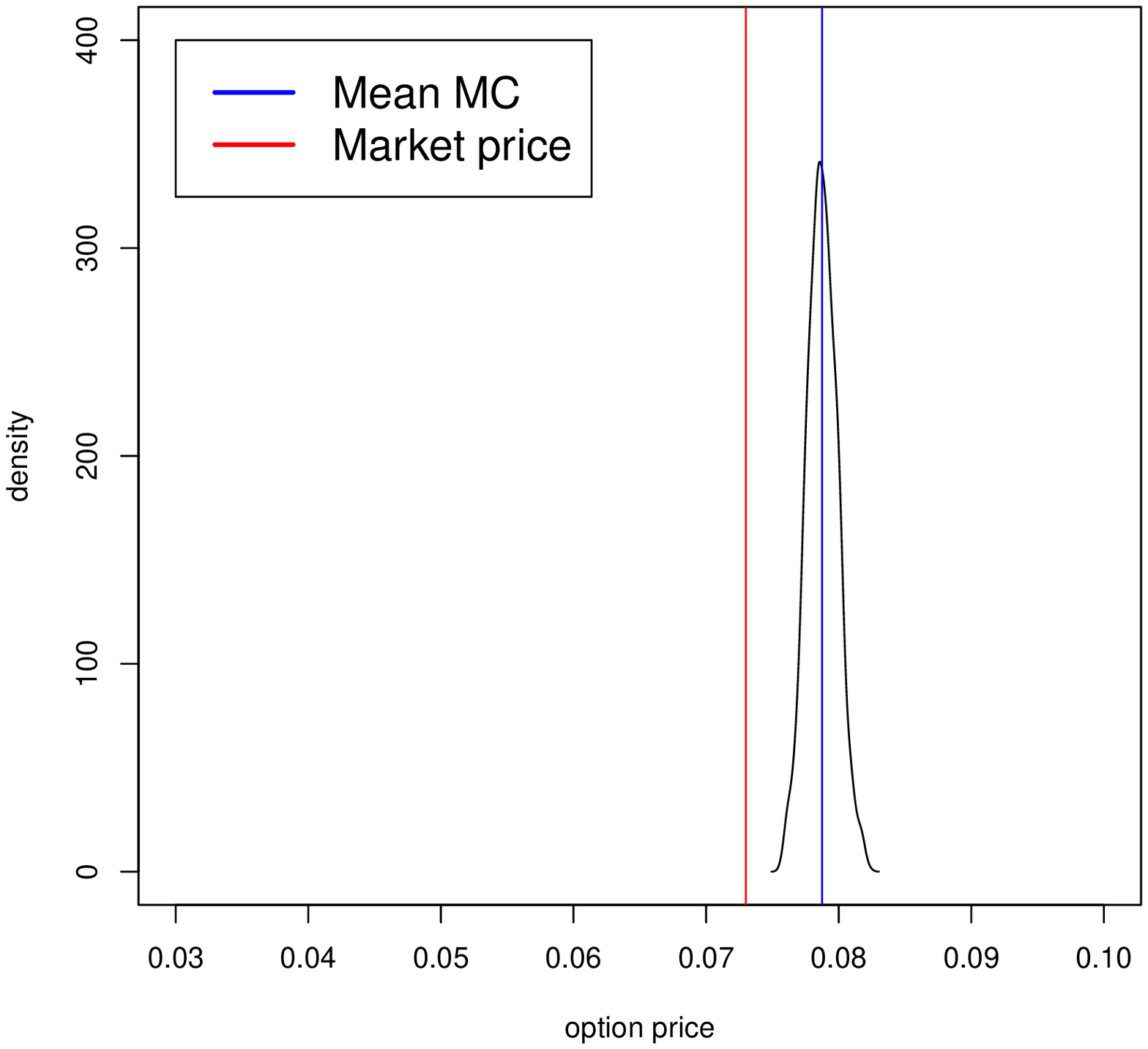}
\end{center}
\vspace{-20pt}
\caption{The smoothed densities of the simulated prices of the put (left) and the call (right) option on December 20, 2011, with strike price 99.50. The distribution is based on 1000  Monte Carlo runs, each of size 10,000.} \label{F:MCprice}
\end{figure}

In this subsection we present detailed estimation results for the date December 20, 2011. A similar procedure has been also conducted for all remaining days under consideration. Assuming the Brace-Gatarek-Musiela dynamics and taking into account the Eurodollar futures closing prices during the period from December 26, 2010 to December 20, 2011 we have conducted PCA and obtained
$$
\left[ \begin{array}{crrr}
0.024063776 & 0.033758193 & 0.040538115 & 0.043033555\\
0.024267981 & 0.018222734 & 0.007111945 & -0.004846372\\
0.007801289 & -0.001039692 & -0.006052515 & -0.004629562
\end{array} \right]$$
as the estimate of $\Lambda$.
To price several put and call options with different strike prices and the closing date falling on December 20, 2011, we have generated 1000 Monte Carlo simulations of size 10,000 and using the least squares algorithm we have obtained estimated option prices for different strike prices. The results are presented in Table~\ref{T:EDprices2}. The Monte Carlo distributions of the prices of the Eurodollar put and call options with the strike price $99.50$ can be seen in Figure~\ref{F:MCprice}. Figure~\ref{F:MCexample} shows examples of 100 Monte Carlo paths, together with the actual realization of the process.

\begin{figure}[!ht]
\begin{center}
\includegraphics[width=6cm]{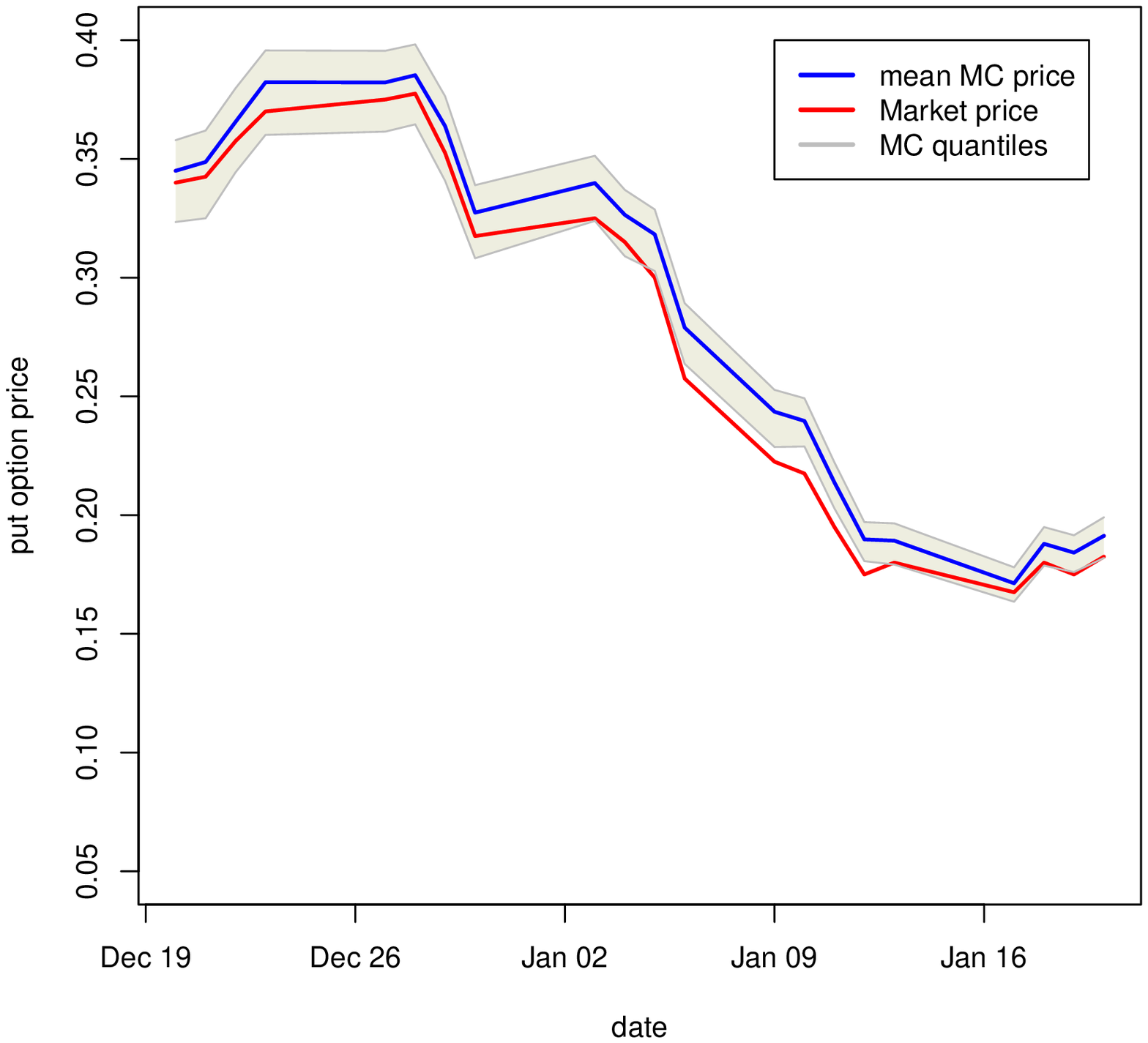}
\includegraphics[width=6cm]{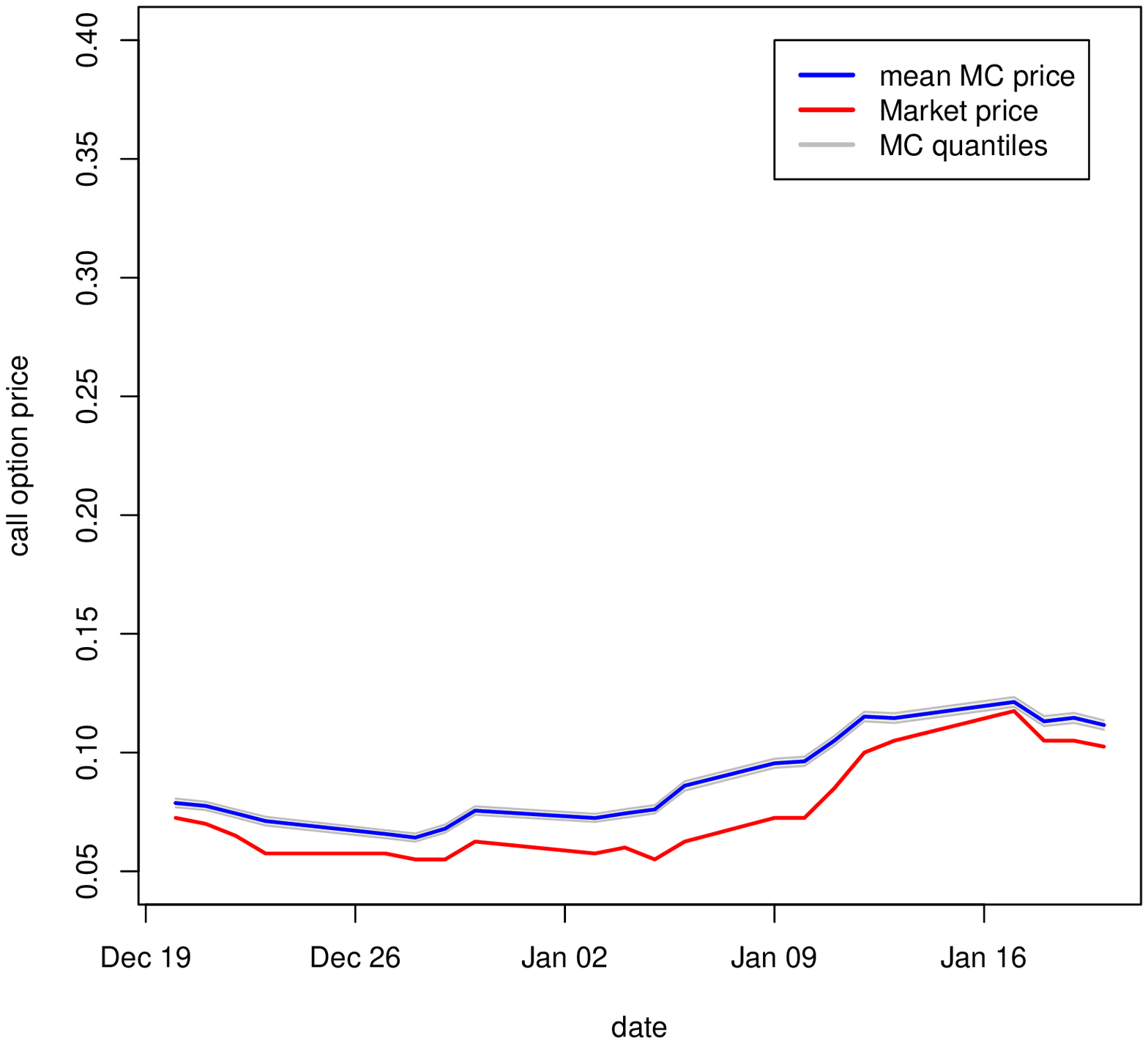}
\end{center}
\vspace{-20pt}
\caption{The estimated prices (with confidence level 90\%) and historical prices of the put (left) and call (right) option prices with strike price 99.50, during the period from December 20, 2011 to January 20, 2012.} \label{F:MCprice2}
\end{figure}

Similar analysis has been performed for all days from December 21, 2011 to January 20, 2012. During that period, EDZ2 was the fourth closest to delivery Eurodollar Future. In Fig.~\ref{F:MCprice2} we can see the dynamics of the original put and call option prices, the sample means from 1000 simulations (each of size 10,000) and the lower and upper 5\% quantiles for the put and call options with strike price $99.50$. The values of the mean and standard deviation of the simulated prices of the options, as well as the corresponding market prices of the options, can be seen in Table~\ref{T:EDprices}. We have chosen the strike price 99.50 because the mean volume of transactions was highest in the considered period. It is interesting to note that the estimated prices corresponding to this strike price stay consistently higher than the market price (see Table~\ref{T:EDprices}), which might be the result of the fact that the option was particularly actively traded (see Table~\ref{T:EDprices2}, where the market price is in most cases lower than the estimated price). It should also be noted that the value of $\sigma$ in Table~\ref{T:EDprices2} is highest for the strike price equal to 99.50, which may explain the interest in the option with this particular strike price.

\begin{table}[!ht]
\caption{The estimated prices of the Eurodollar options on December 20, 2011, based on 1000 simulations (each of size 10,000). Here $\mu$ denotes the sample mean of the 1000 prices obtained with MC simulation, while $\sigma$ denotes the sample standard deviation.}

\bigskip
\begin{center}
\begin{tabular}{|c|ccc|ccc|}
\hline
Date: Dec. 20, 2011& \multicolumn{3}{|c|}{Put} & \multicolumn{3}{|c|}{Call}\\
\hline
 Strike price & Market price &$\mu$ & $\sigma$ & Market price & $\mu$ & $\sigma$\\
\hline
98.00 & 0.070 & 0.045 & 0.0038 & 1.295 & 1.267 & 0.0038 \\ 
98.12 & 0.078 & 0.052 & 0.0043 & 1.178 & 1.154 & 0.0043 \\ 
98.25 & 0.085 & 0.061 & 0.0044 & 1.060 & 1.032 & 0.0044 \\ 
98.37 & 0.095 & 0.070 & 0.0048 & 0.945 & 0.922 & 0.0048 \\ 
98.50 & 0.105 & 0.082 & 0.0050 & 0.833 & 0.804 & 0.0050 \\ 
98.62 & 0.120 & 0.096 & 0.0055 & 0.723 & 0.698 & 0.0055 \\ 
98.75 & 0.138 & 0.114 & 0.0058 & 0.615 & 0.587 & 0.0058 \\ 
98.87 & 0.155 & 0.134 & 0.0061 & 0.508 & 0.488 & 0.0061 \\ 
99.00 & 0.175 & 0.160 & 0.0067 & 0.403 & 0.386 & 0.0067 \\ 
99.12 & 0.203 & 0.191 & 0.0078 & 0.308 & 0.298 & 0.0078 \\ 
99.25 & 0.238 & 0.232 & 0.0088 & 0.218 & 0.211 & 0.0088 \\ 
99.37 & 0.280 & 0.280 & 0.0097 & 0.135 & 0.141 & 0.0097 \\ 
99.50 & 0.340 & 0.345 & 0.0110 & 0.073 & 0.079 & 0.0110 \\ 
99.62 & 0.425 & 0.421 & 0.0094 & 0.033 & 0.036 & 0.0094 \\ 
99.75 & 0.525 & 0.528 & 0.0044 & 0.008 & 0.009 & 0.0044 \\ 
\hline
\end{tabular}
\end{center}
\label{T:EDprices2}
\end{table}

\begin{table}[!ht]
\caption{Simulated and historical prices of the Eurodollar options with strike price 99.50 based on 1000 Monte Carlo simulations (each of size 10,000). As before, $\mu$ denotes the sample mean, whereas $\sigma$ denotes the sample standard deviation of the prices calculated form the simulations.}

\bigskip
\begin{center}
\begin{tabular}{|c|ccc|ccc|}
\hline
\multirow{2}{*}{Date} & \multicolumn{3}{|c|}{Put} & \multicolumn{3}{|c|}{Call}\\
\cline{2-7}
 & Market price & $\mu$ & $\sigma$ & Market price & $\mu$ &  $\sigma$\\
\hline
Dec.20, 2011 & 0.340 & 0.345 & 0.0110 & 0.073 & 0.079 & 0.0011 \\ 
Dec.21, 2011 & 0.342 & 0.349 & 0.0110 & 0.070 & 0.078 & 0.0011 \\ 
Dec.22, 2011 & 0.358 & 0.366 & 0.0112 & 0.065 & 0.074 & 0.0011 \\ 
Dec.23, 2011 & 0.370 & 0.382 & 0.0111 & 0.058 & 0.071 & 0.0011 \\ 
Dec.27, 2011 & 0.375 & 0.382 & 0.0103 & 0.058 & 0.066 & 0.0010 \\ 
Dec.28, 2011 & 0.378 & 0.385 & 0.0102 & 0.055 & 0.064 & 0.0010 \\ 
Dec.29, 2011 & 0.352 & 0.364 & 0.0104 & 0.055 & 0.068 & 0.0010 \\ 
Dec.30, 2011 & 0.318 & 0.327 & 0.0093 & 0.062 & 0.076 & 0.0011 \\ 
Jan.03, 2012 & 0.325 & 0.340 & 0.0088 & 0.058 & 0.072 & 0.0011 \\ 
Jan.04, 2012 & 0.315 & 0.326 & 0.0086 & 0.060 & 0.074 & 0.0011 \\ 
Jan.05, 2012 & 0.300 & 0.318 & 0.0084 & 0.055 & 0.076 & 0.0011 \\ 
Jan.06, 2012 & 0.258 & 0.279 & 0.0079 & 0.062 & 0.086 & 0.0012 \\ 
Jan.09, 2012 & 0.222 & 0.244 & 0.0072 & 0.072 & 0.095 & 0.0012 \\ 
Jan.10, 2012 & 0.218 & 0.240 & 0.0071 & 0.072 & 0.096 & 0.0012 \\ 
Jan.11, 2012 & 0.195 & 0.214 & 0.0069 & 0.085 & 0.105 & 0.0012 \\ 
Jan.12, 2012 & 0.175 & 0.190 & 0.0054 & 0.100 & 0.115 & 0.0013 \\ 
Jan.13, 2012 & 0.180 & 0.189 & 0.0054 & 0.105 & 0.115 & 0.0013 \\ 
Jan.17, 2012 & 0.168 & 0.171 & 0.0046 & 0.118 & 0.121 & 0.0012 \\ 
Jan.18, 2012 & 0.180 & 0.188 & 0.0053 & 0.105 & 0.113 & 0.0013 \\ 
Jan.19, 2012 & 0.175 & 0.184 & 0.0053 & 0.105 & 0.115 & 0.0013 \\ 
Jan.20, 2012 & 0.182 & 0.191 & 0.0054 & 0.102 & 0.112 & 0.0012 \\ 
\hline
\end{tabular}
\end{center}
\label{T:EDprices}
\end{table}


\subsection{Basket and dual-strike options}
In this subsection we will use the least squares algorithm to price 1.5 month basket and dual-strike American put options whose payoff functions are based on two market indices, namely DAX and EUROSTOXX50. The latter will be denoted by the symbol EUR for brevity. We will assume that the underlying instruments follow the standard bivariate Brownian dynamics. Unfortunately, bivariate options are usually over-the-counter (OTC) instruments, so it is difficult to find market data for such options. Nevertheless, we could do a partial comparison with the relevant one-dimensional standard American put options based on DAX and EUR. As was the case with the previous example, we start with some background information.

\subsubsection{DAX and EUROSTOXX50 Indices.}
The univariate standard American put options based on DAX and EUR are traded on the Eurex Exchange. In fact, the underlyings are not indices but exchange-traded funds (ETF), which are actively traded on the German stock market (Deutsche B{\"o}rse Group). The DAX and EUR indices are highly correlated, chiefly due to the inclusion of some common stocks in their baskets. The estimated value of Pearson's linear correlation coefficient for the period from October 23, 2012 to January 08, 2013 is equal to 0.920. Some contagion between these indices might potentially occur, but in such a short period of time this aspect is negligible. In general, the issue of contagion could be addressed by adopting models with different dynamics (e.g. of the multivariate GARCH variety). Such approach would be very closely related to the Heston and Nandi option pricing model~\cite{HesNan2000}, which is the methodology we will adopt in the last example.

\subsubsection{Basket and dual-strike options.}
As has been already stated, basket and dual-strike options are mainly OTC derivatives. In this example we will consider a bivariate American put option. The payoff functions at time $t$, for a bivariate basket American put option (1) and dual-strike American put option (2) is given by
$$p^{(1)}(t)=\max\big(K_{1}-S_{1}(t),K_{2}-S_{2}(t),0\big),\quad p^{(2)}(t)=\max\left(\frac{K_{1}+K_{2}}{2}-\frac{S_{1}(t)+S_{2}(t)}{2},0\right),$$
where $S_{1}(t)$ and $S_{2}(t)$ are the prices of the first and the second underlying at time $t$, respectively, and $K_{1}$, $K_{2}$ are the strike prices.

\subsubsection{Model setup, data details and implementation parameters.}
We will be assuming that the price process $(S_{1}(t),S_{2}(t))$ is modeled by a 2-dimensional geometric Brownian motion, with the instantaneous correlation coefficient and instantaneous standard deviations for the processes $\log S_1$ and $\log S_2$ denoted by $\boldsymbol{\rho},\boldsymbol{\sigma}_1$ and $\boldsymbol{\sigma}_2$, respectively. 
We will construct a bivariate basket and dual-strike American put options based on 1 DAX ETF share and 2.5 EUR ETF shares (to have similar strike prices in both cases). We will price Basket and Dual-strike put options on January 08, 2013 with the expiration date March 16, 2013 (to make it comparable to existing univariate options). The option lifetime will be 49 business days. The strike prices will range from 65 to 75 and from 66 to 76, for the first and the second strike price, respectively.

To estimate $\boldsymbol{\rho},\boldsymbol{\sigma}_1$ and $\boldsymbol{\sigma}_2$ we will use the last 50 observations of the price of ETF (DE) DAX and ETF (DE) EUROSTOXX50. Choosing a relatively short time interval for calibration purposes is quite common in practice  (e.g. this is the case with the estimation of the VIX volatility index). 

As in the previous case, we will need two inputs for the least squares algorithm: an interest rate (for discounting) and appropriate basis functions. Because the option lifetime is short, we will assume that the interest rate is constant and equal to $r=1.50\%$ (the ECB interest rate on January 08, 2013). Moreover, we will use the following exponentially weighted polynomials of two variables to perform the regression:
$$e^{\frac{1}{2}},\quad e^{\frac{-x}{2}}x,\quad e^{\frac{-y}{2}}y,\quad e^{\frac{-(x+y)}{4}}xy,
\quad e^{\frac{-(x+y)}{4}}xy^{2},\quad e^{\frac{-(x+y)}{4}}x^{2}y,\quad e^{\frac{-(x+y)}{4}}x^{2}y^{2}.
$$

\subsubsection{Estimation and numerical results.}
The estimated (annualized) covariance matrix gives us the values $\boldsymbol{\rho}=0.920$, $\boldsymbol{\sigma}_{1}= 0.133$ and $\boldsymbol{\sigma}_{2}=0.119$. Using these numbers, we run 100,000 Monte Carlo simulations (each of size 49). Next, using the least squares algorithm we compute the prices of the basket and dual-strike American put options for different strike prices. We also compute the least squares prices for the univariate American put options based on 1 DAX ETF share and 2.5 EUR ETF share. Apart from the market data, we also present the theoretical price according to the Cox-Ross-Rubinstein model (CRR) as it is used by the Eurex Exchange to quote option prices when no trading takes place. It should be noted that the volume of transaction of American put options is very low, so unfortunately the market price is just for comparison purposes. Also, the least squares price should be compared with the CRR price rather than the market price (as it is computed under the compatible assumptions about the asset dynamics).

\begin{figure}[!ht]
\begin{center}
\includegraphics[width=6cm]{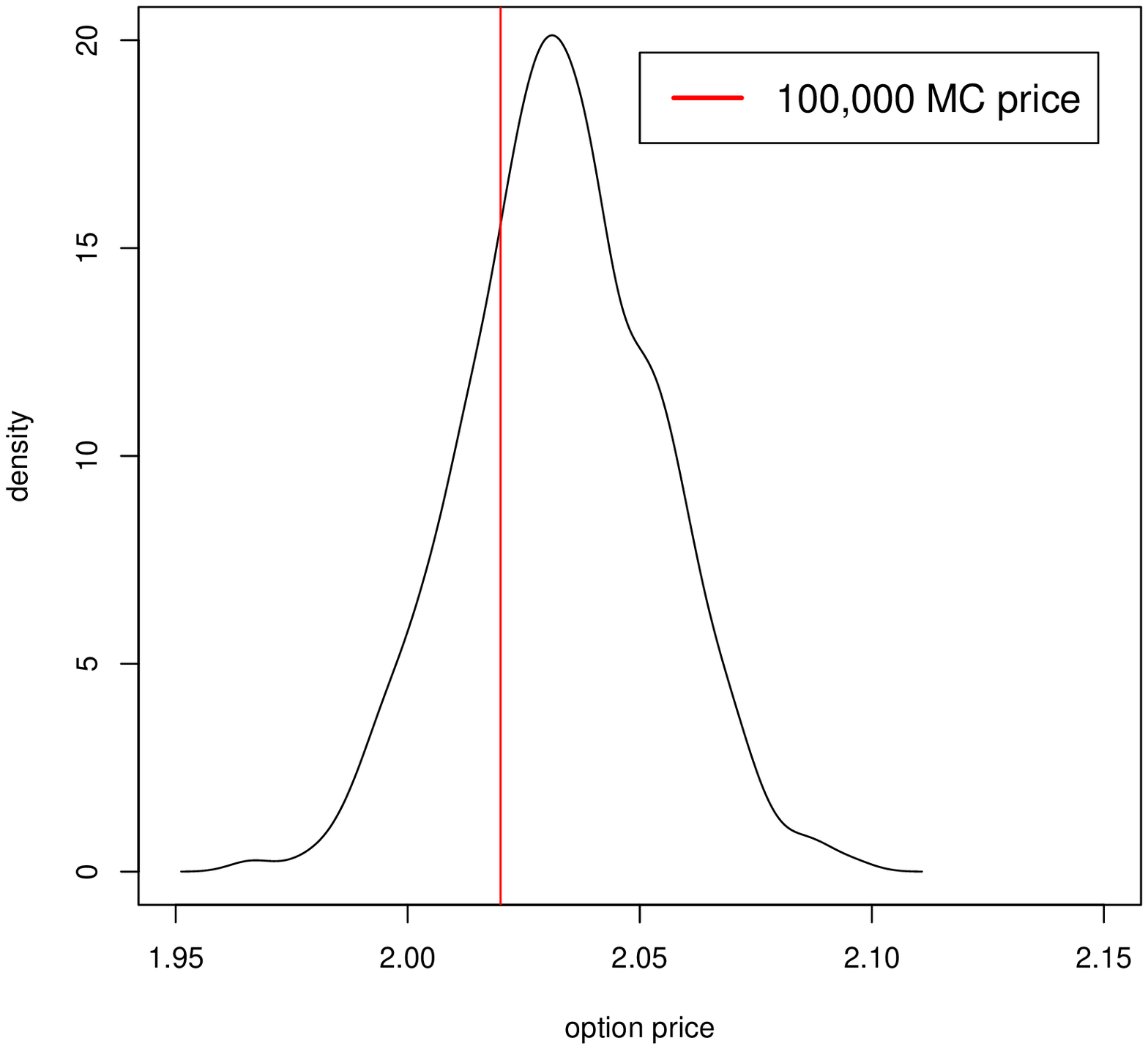}
\includegraphics[width=6cm]{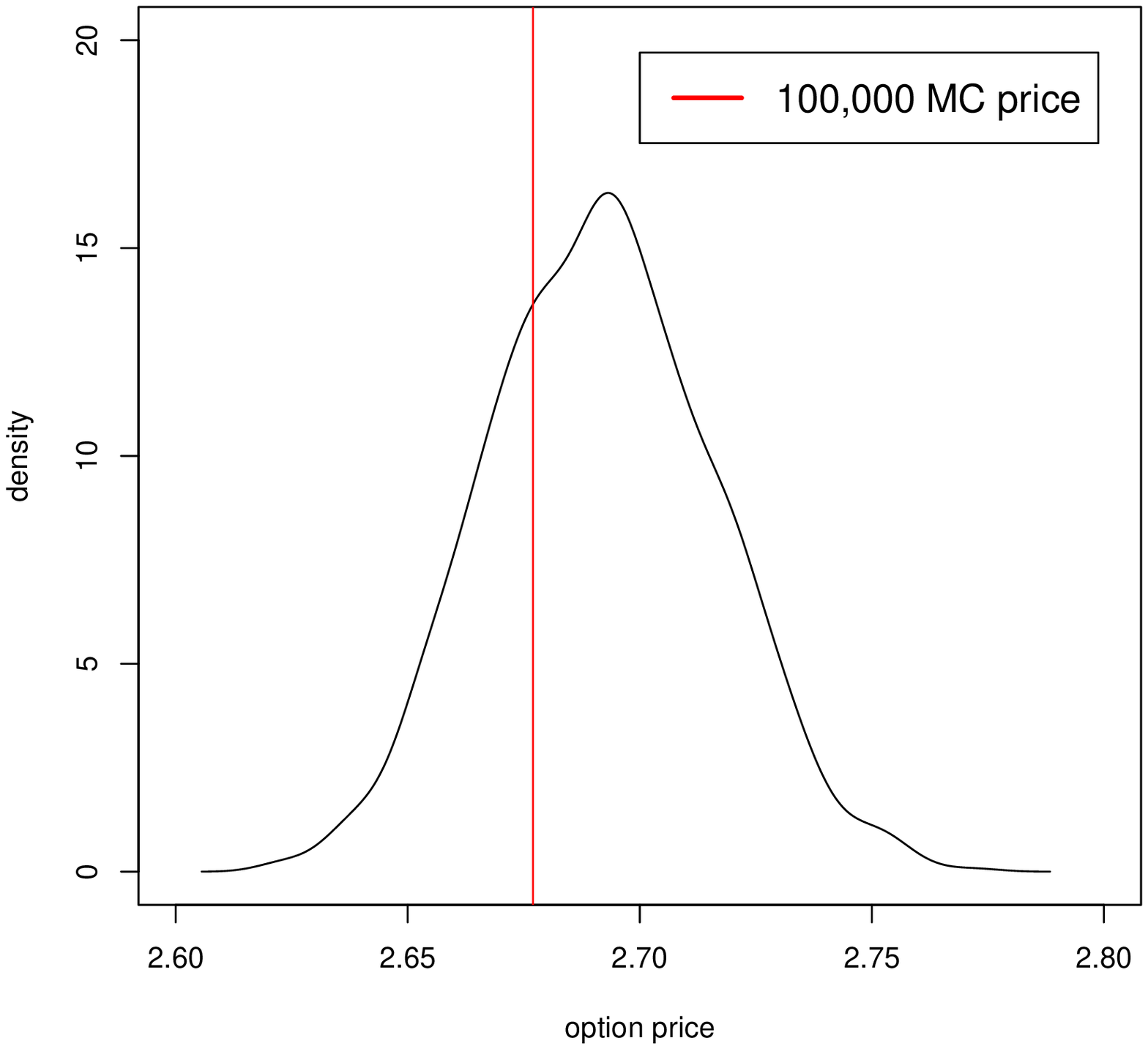}
\end{center}
\caption{The smoothed densities of the least squares prices of the basket (left) and the dual-strike (right) American put options for the strike prices $K_1=K_2=70$. The vertical lines depict the sample mean of the least squares prices.} \label{F:optionBASKET}
\end{figure}

The prices (obtained using single 100,000 Monte Carlo run) can be seen in Table~\ref{T:OPTIONprices}. The columns with names DAX and EUR denote the standard univariate put options (i.e. with 1 DAX ETF and 2.5 EUR ETF share as the underlying, respectively). We have also performed multiple Monte Carlo runs (1000), each of size 10,000 for the basket and dual-strike options with the strike prices $K_1=K_2=70$. The corresponding Monte Carlo density function could be seen in Figure~\ref{F:optionBASKET} (this could provide some information about the model and/or the Monte Carlo bias).

\begin{table}[!ht]
\caption{Prices of the options according to historical stock market data, the CRR model and the 
least squares algorithm. Here $S_0=(68.05,69.72)$, $r=1.50\%$, $T=49/252$, $\boldsymbol{\sigma}_{1}=0.133$, $\boldsymbol{\sigma}_{2}=0.119$, $\boldsymbol{\rho}=0.920$.}
\bigskip
\begin{center}
\begin{tabular}{|cc|cc|cc|cccc|}
\hline  
\multicolumn{2}{|c|}{Strike price}  &  \multicolumn{2}{|c|}{Market price} & \multicolumn{2}{|c|}{CRR price} & \multicolumn{4}{|c|}{least squares price}\\
\hline
EUR & DAX & EUR & DAX & EUR & DAX & EUR & DAX & Basket & Dual-Strike\\ 
\hline
65.0 & 66 & 0.58 & 0.34 & 0.45 & 0.25 & 0.44 & 0.24 & 0.31 & 0.46 \\ 
67.5 & 66 & 1.50 & 0.34 & 1.25 & 0.25 & 1.23 & 0.24 & 0.59 & 1.23 \\ 
70.0 & 66 & 3.10 & 0.34 & 2.67 & 0.25 & 2.63 & 0.24 & 1.00 & 2.65 \\ 
72.5 & 66 & 5.15 & 0.34 & 4.63 & 0.25 & 4.62 & 0.24 & 1.58 & 4.62 \\ 
75.0 & 66 & 7.53 & 0.34 & 6.95 & 0.25 & 6.95 & 0.24 & 2.33 & 6.95 \\ 
\hline
65.0 & 68 & 0.58 & 0.85 & 0.45 & 0.69 & 0.44 & 0.67 & 0.52 & 0.72 \\ 
67.5 & 68 & 1.50 & 0.85 & 1.25 & 0.69 & 1.23 & 0.67 & 0.91 & 1.27 \\ 
70.0 & 68 & 3.10 & 0.85 & 2.67 & 0.69 & 2.63 & 0.67 & 1.45 & 2.65 \\ 
72.5 & 68 & 5.15 & 0.85 & 4.63 & 0.69 & 4.62 & 0.67 & 2.17 & 4.62 \\ 
75.0 & 68 & 7.53 & 0.85 & 6.95 & 0.69 & 6.95 & 0.67 & 3.04 & 6.95 \\ 
\hline
65.0 & 70 & 0.58 & 1.74 & 0.45 & 1.52 & 0.44 & 1.49 & 0.82 & 1.50 \\ 
67.5 & 70 & 1.50 & 1.74 & 1.25 & 1.52 & 1.23 & 1.49 & 1.33 & 1.64 \\ 
70.0 & 70 & 3.10 & 1.74 & 2.67 & 1.52 & 2.63 & 1.49 & 2.01 & 2.67 \\ 
72.5 & 70 & 5.15 & 1.74 & 4.63 & 1.52 & 4.62 & 1.49 & 2.86 & 4.62 \\ 
75.0 & 70 & 7.53 & 1.74 & 6.95 & 1.52 & 6.95 & 1.49 & 3.85 & 6.95 \\ 
\hline
65.0 & 72 & 0.58 & 3.00 & 0.45 & 2.79 & 0.44 & 2.75 & 1.22 & 2.76 \\ 
67.5 & 72 & 1.50 & 3.00 & 1.25 & 2.79 & 1.23 & 2.75 & 1.86 & 2.77 \\ 
70.0 & 72 & 3.10 & 3.00 & 2.67 & 2.79 & 2.63 & 2.75 & 2.67 & 3.09 \\ 
72.5 & 72 & 5.15 & 3.00 & 4.63 & 2.79 & 4.62 & 2.75 & 3.64 & 4.63 \\ 
75.0 & 72 & 7.53 & 3.00 & 6.95 & 2.79 & 6.95 & 2.75 & 4.72 & 6.95 \\ 
\hline
65.0 & 76 & 0.58 & 6.43 & 0.45 & 6.29 & 0.44 & 6.28 & 2.33 & 6.28 \\ 
67.5 & 76 & 1.50 & 6.43 & 1.25 & 6.29 & 1.23 & 6.28 & 3.24 & 6.28 \\ 
70.0 & 76 & 3.10 & 6.43 & 2.67 & 6.29 & 2.63 & 6.28 & 4.28 & 6.28 \\ 
72.5 & 76 & 5.15 & 6.43 & 4.63 & 6.29 & 4.62 & 6.28 & 5.41 & 6.30 \\ 
75.0 & 76 & 7.53 & 6.43 & 6.95 & 6.29 & 6.95 & 6.28 & 6.62 & 7.16 \\ 

\hline
\end{tabular}
\end{center}
\label{T:OPTIONprices}
\end{table}

\subsection{The Heston-Nandi model}
In the last example we will use the least squares algorithm to price two 1.5 month American put options whose payoff is based on a single market index. We will use data from the previous example, i.e., we will price options written on DAX and EUR indices. We will assume that the dynamics of the underlying instruments could be described with the Heston-Nandi GARCH model~\cite{HesNan2000}.

Let $S_t$ denote the price of the underlying. Using the Heston-Nandi GARCH dynamics, we assume that the log-returns of the random process $S_t$ could be described by formula
$$\Delta\log S_t=r_{\textrm{daily}}+\lambda\sigma_{t}^{2}+\sigma_{t}\epsilon_{t},$$
with
$\sigma_{t}^{2}=\omega+\beta\sigma_{t-1}^{2}+\alpha(\epsilon_{t-1}-\gamma\sigma_{t-1})^{2},$
where $\Delta$ denotes the daily backward difference, the parameter $r_{\textrm{daily}}$ denotes daily risk-free interest rate, $(\lambda, \omega, \beta, \alpha, \gamma)$ are model parameters and $\epsilon_{t}$ is the standard Gaussian white noise. In addition, we will assume that there is no asymmetry in the model, i.e. $\gamma=0$.

If we use the standard Heston-Nandi dynamics (with the objective probability measure) then the discounting part of the least squares algorithm will be path dependent. In order to avoid this complication we will switch to the risk-neutral measure and use the risk-neutral dynamic of the underlying return. The risk neutral process is obtained simply by replacing (previously estimated) parameters $\lambda$ and $\gamma$ with $(-0.5)$ and $(\gamma+\lambda+0.5)$, respectively (see~\cite{HesNan2000} for details). Moreover, we will use the long run expected standard deviation from Heston Nandi model for comparison purposes (see~\cite{HesNan2000}):
\begin{equation}\label{hn.sd}
\sigma_{HN}=\frac{\omega+\alpha}{1-\beta-\alpha\gamma^{2}}.
\end{equation}
The EUR and DAX data will be used again as the underlyings. As before, the options expiration date will be March 16, 2013 and we will price them on January 08, 2013 (thus the option lifetime $T$ will be 49 business days). The weighted Laguerre polynomials of degree not greater than 3 will serve
as the base polynomials for the regression procedure. 

As before, we will assume that (annualised) risk free rate is equal to $r=1.50\%$ and put $r_{\textrm{daily}}=r/252$ (as there are 252 trading days each year).
Using the last 50 prices of 2.5 EUR ETF and 1 DAX ETF shares, we have obtained two sets of parameters:

\begin{center}
\def\arraystretch{1.2}
\begin{tabular}{|c|c|c|c|c|}
\hline
&$\lambda$&$\omega$&$\alpha$&$\beta$\\
\hline
EUR&7.280&2.738$\times10^{-5}$&5.238$\times10^{-5}$&0.086\\
\hline
DAX&16.971&1.954$\times10^{-5}$&5.404$\times10^{-5}$&4.758$\times10^{-28}$\\
\hline
\end{tabular}
\end{center}

\begin{figure}[!ht]
\begin{center}
\includegraphics[width=6cm]{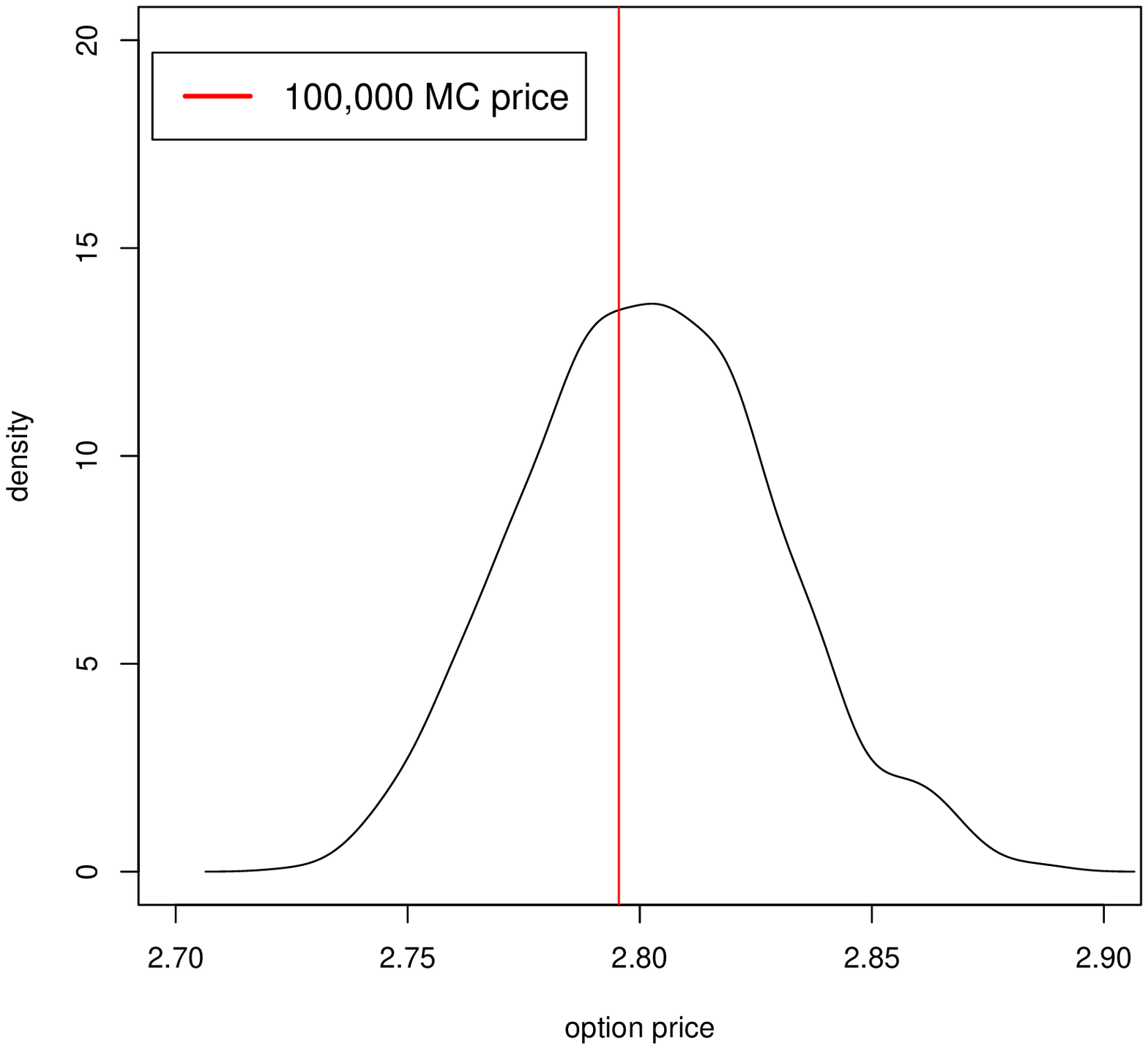}
\includegraphics[width=6cm]{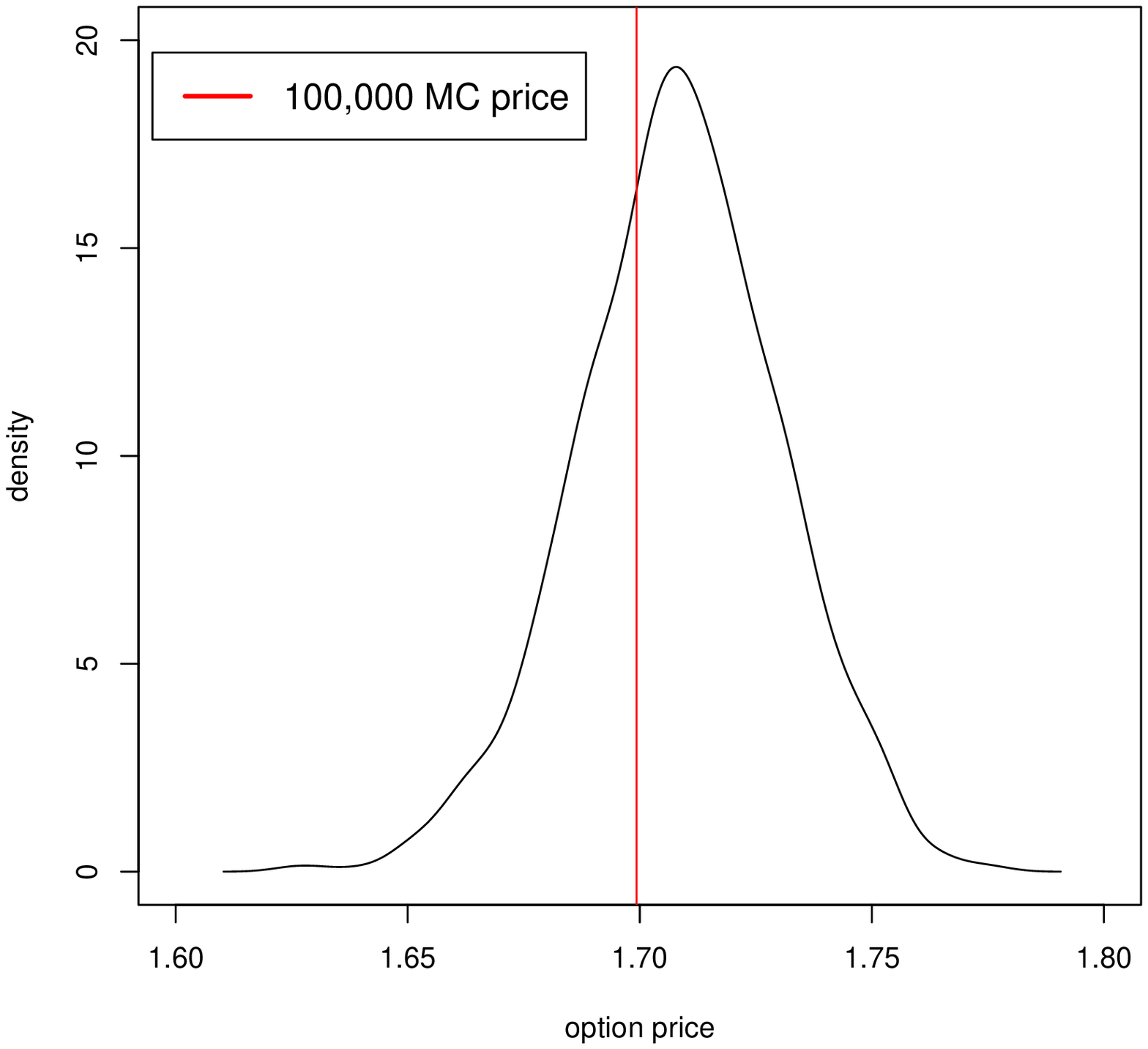}
\end{center}
\caption{The smoothed distributions of the least squares prices of the EUR (left) and DAX (right) American put options. The vertical lines correspond to the sample mean of the least squares prices.}\label{F:optionHN}
\end{figure}

The initial values of the underlying are 68.05 and 69.72, respectively, in the EUR and DAX case. The (annualized) volatilities obtained from~\eqref{hn.sd} are equal to $0.149$ and $0.137$, respectively. The mean sample prices of American put options obtained from ten simulations (each consisting of 100,000 Monte Carlo paths) can be seen in Table~\ref{T:EUROSTOXX_HN}. We also present the theoretical European put option prices according to Heston-Nandi model~\cite{HesNan2000}, as well as the American put options prices and early exercise premiums (i.e. the differences between the prices of American and European put options) according to Cox-Ross-Rubinstein model (CRR), with volatilities obtained from~\eqref{hn.sd}. Both models are presented for comparison purposes. Moreover, we perform multiple Monte Carlo runs (1000), each of size 10,000, to calculate prices of the American put options with the strike price 70 (both for EUR and DAX). Smoothed simulated probability density functions are plotted in Figure~\ref{F:optionHN}.

\begin{table}[!ht]
\caption{Prices of the EUR and DAX American put options according to the least squares algorithm (L-S), compared with the actual market prices, CRR model prices and the Heston-Nandi European put option prices. EA denotes the early exercise premium.}

\bigskip
\begin{center}
\begin{tabular}{|c|c|cc|cc|}
\hline
\multicolumn{6}{|c|}{EUR American put options}\\
\hline  
Strike price  & Market price & CRR price & CRR EA & H-N price & L-S price\\
\hline
65.0 & 0.58 & 0.57 & 0.00 & 0.57 & 0.57\\
67.5 & 1.50 & 1.41 & 0.01 & 1.40 & 1.40\\
70.0 & 3.10 & 2.79 & 0.03 & 2.78 & 2.79\\
72.5 & 5.15 & 4.67 & 0.06 & 4.66 & 4.71\\
75.0 & 7.53 & 6.88 & 0.11 & 6.88 & 6.98\\
\hline
\multicolumn{6}{|c|}{DAX American put options}\\
\hline  
Strike price  & Market price & CRR price & CRR EA & H-N price & L-S price\\
\hline
66 & 0.34 & 0.38 & 0.00 & 0.38 & 0.38\\
68 & 0.85 & 0.88 & 0.01 & 0.87 & 0.87\\
70 & 1.74 & 1.74 & 0.01 & 1.70 & 1.70\\
72 & 3.00 & 2.98 & 0.03 & 2.91 & 2.92\\
76 & 6.43 & 6.33 & 0.10 & 6.22 & 6.31\\
\hline
\end{tabular}
\end{center}
\label{T:EUROSTOXX_HN}
\end{table}

\section{Concluding remarks}
We have shown that the widely used least squares approach to Monte Carlo based pricing of American options remains valid under very general and flexible choice of assumptions. In particular,  convergence to the theoretical price obtained via Snell envelopes remains true with a highly adaptable setup for approximation of conditional expectations. Of course one should be aware that the computational cost of liberalization of the assumptions may be potentially very high. However, a growing body of empirical evidence indicates that in many practical applications even relatively limited non-linear extensions of standard regression may produce satisfactory results, as illustrated also by our three examples. The relaxation of the assumptions of the method should be seen primarily as increase in freedom of choice of settings for a specific implementation of the algorithm, which with careful choices may nevertheless retain computational viability.

\subsubsection*{Acknowledgments:}
The second author acknowledges the support by Project operated within the Foundation for Polish Science IPP Programme "Geometry and Topology in Physical Models" co-financed by the EU European Regional Development Fund, Operational Program Innovative Economy 2007-2013.

\bibliographystyle{plainnat}
\bibliography{bibliografia}

\end{document}